%% file: main.tex
\documentclass[authoryear]{elsarticle}

\usepackage{cleveref}
\usepackage{booktabs}
\usepackage{subfig}

\begin{document}

\begin{frontmatter}



\title{Situated Action in Pre-Hospital Critical Care Dispatch: Identifying where and how Algorithmic Assistance might be useful in the daily work of specialist Emergency Medical Dispatchers} %% Article title

\affiliation[1]{organization={Swansea University},
            addressline={Computational Foundry, Fabian Way},
            city={Swansea},
            postcode={SA1 8EN},
            state={Wales},
            country={UK}}

\affiliation[2]{organization={EMRTS, Swansea Bay UHB},
            addressline={Ffordd Angel},
            city={Llanelli},
            postcode={SA14 8LQ},
            state={Wales},
            country={UK}}

\author[1]{Ben Wilson\corref{cor1}} 
\ead{b.j.m.wilson@swansea.ac.uk}
\author[1]{Matt Roach}
\author[2]{Greg Browning}
\author[2]{Chris Connor}
\author[2]{David Rawlinson}

\cortext[cor1]{Corresponding author}

%% Abstract
\begin{abstract}
This study uses ethnographic immersion and observation as contextual inquiry to understand situated action at an Emergency Medical Dispatch critical care hub. 
The work is a response to the urgent need to recruit context-specific knowledge and participation into design that helps to narrow the \textit{AI Chasm} - the gap between the promise of Artificial Intelligence (AI) systems and what they deliver for clinicians and their patients.

The work of a pre-hospital critical care team's dispatch process is described and analysed to reveal both the structure of the workflow and the different cognitive demands it makes on staff tasked with dispatch decision-making. 
Elements of attention, communication and focus between the humans, as they carry out this work, are drawn out in order to understand the work-as-done and identify the many dependencies in the process.

The motivation is to establish where AI support might be useful and to discover what challenges there could be in designing appropriate algorithmic assistance. 
We ask whether, where and how the design and implementation of an AI system might be considered. 
The ultimate objective is to improve the decision process itself to the benefit of clinicians and patients. 

The study identifies three key steps in the situated workflow and details how decision-makers negotiate each one as emergency calls follow complex routes between them. 
We find compelling evidence that the first of these decision steps constitutes the most promising candidate for unobtrusive assistance that could be safe and effective in improving both clinician workload and clinical outcomes.
\end{abstract}


% \begin{highlights}
% \end{highlights}

%% Keywords
\begin{keyword}


% MSC codes here, in the form: \MSC code \sep code
% or \MSC[2008] code \sep code (2000 is the default)

Human Computer Interaction \sep 
Ethnography \sep 
Observation \sep 
Contextual Design \sep
Situated Action \sep
Clinical Decision Support
\end{keyword}

\end{frontmatter}



\input{incl_01_intro}

\input{incl_02_methods}
\input{incl_03_results}
\input{incl_04_discussion}

\input{incl_05_conclusions}

\setlength{\parindent}{0pt}
\input{incl_99_declarations}

% \appendix
% \section{Example Appendix Section}


\bibliographystyle{elsarticle-harv} 
\bibliography{citations}

\end{document}

%% file: incl_01_intro.tex
% !TEX root = ../main.tex

\section{Introduction}
    \label{sec:intro}
    
\subsection{Motivation}    
    \label{ssec:intro:motivation}
    
    Despite advances in Artificial Intelligence (AI) models, a recognised problem remains one of translation from the AI lab bench to the clinic.
    This \textit{AI chasm}~\citep{keane2018EyeAIAutonomousa,cabitza2020BridgingLastMilea,vasey2021decide,marwaha2022CrossingChasmModel,aristidou2022BridgingChasmAI} is the large gap between the promise of AI and the clinical benefit it actually brings.
    This challenge, in moving from bench to bedside - or more generally \textit{from bench performance to situated performance}~\citep{wilson2025DimensionsHumanMachineCombination}, points toward the need for situated design, development and evaluation to ensure sustainable applicability within real clinical workflows~\citep{kelly2019KeyChallengesDelivering,ganguli2020MachineLearningPursuit,verma2021ImplementingMachineLearninga,zajac2023ClinicianFacingAIWild}. 
    An experienced team at a US academic medical centre urge against a narrow focus on bench performance, emphasising that such improvements alone are insufficient for clinical benefit.
    They argue that developers must understand the complexities of situated care delivery \textit{`before building the ML model'}~\cite[p1]{li2020DevelopingDeliveryScience}.

    Ethnography and Contextual Inquiry are fundamental parts of Human Computer Interaction (HCI) studies~\citep{wixon1994ContextualInquiryGrounding,beyer1999ContextualDesign,gulliksen2003KeyPrinciplesUsercentred}.
    But they have been less prominent in studies of human-AI combination, especially those focusing on clinical contexts.
    This has led to issues becoming evident in deployment, such as relevance, neglible impact, non-adoption, aversion or even the generation of hostility~\citep{schwartz2021ClinicianInvolvementResearch,marwaha2022CrossingChasmModel,aristidou2022BridgingChasmAI,panigutti2022UnderstandingImpactExplanations}.
    At the same time, several writers emphasise that role-specific needs and sociotechnical context are important factors that must be taken into account~\citep{verma2021ImplementingMachineLearninga,panigutti2022UnderstandingImpactExplanations,zajac2023ClinicianFacingAIWild}.
    Without an appreciation of context, algorithmic tool designs frequently fail to translate into real world benefit~\citep{keane2018EyeAIAutonomousa,cabitza2019ProofPuddingPraise}.
    We demonstrate that context is critical for AI development to consider, since it can determine fundamental questions about the earliest of decisions.
    Whether to develop at all. If so, for where in the workflow. And how assistance should aim to fit into that point of the workflow, not only as it is now but also in the future -- given how the sociotechnical context can be expected to evolve.

\subsection{Context}
        \label{ssec:intro:context_em_med_dispatch} 

    Nearly 2,000 emergency medical calls per day come into the clinical control centre of the Welsh Ambulance Service\footnote{Welsh Ambulance Services University NHS Trust} in Cwmbran, UK. 
    Within the control centre, a very small and specialist team of clinicians and allocators from the pre-hospital critical care service\footnote{Known as the Welsh Emergency Medical Retrieval and Transfer Service (EMRTS, Cymru)} play a distinct role.
    They operate the \textit{Critical Care Hub} in which they follow their own complex multi-step process to dispatch critical care teams to the field for a small minority of cases -- an average of 15 incidents per day.
    The decision to task a team to the field is taken once the team at this critical care hub are satisfied that the information from an incident means it is likely a patient will benefit from the highly specialist intervention their critical care colleagues can bring.
            
    The EMRTS pre-hospital critical care service was established in April 2015 to provide rapid pre-hospital critical care across Wales. 
    It has bases operated by an air ambulance charity \footnote{Wales Air Ambulance Charitable Trust (WAACT)} at four locations across Wales -- Caernarfon, Welshpool, Llanelli and Cardiff -- with a control centre, the critical care hub, at a fifth location, Cwmbran. 
    In addition to four aircraft, the service coordinates the activities of a fleet of rapid response road vehicles. 
    The critical care service is consultant-led, with specialists from Emergency Medicine, Anaesthesia and Intensive Care Medicine, while the core of the service is formed by Critical Care Practitioners who deliver pre-hospital critical care in field teams and rotate into the critical care hub itself to provide clinical decision-making in the dispatch process.
    It is decision-making in the critical care hub that forms the immediate context for this work.

    The usual configuration at the critical care hub is a team of two, one of whom is a critical care clinician.
    The other is an experienced allocator.
    Their shifts last for 12 hours - during which they will monitor a dynamic list of all the emergency (999) calls received by the Welsh ambulance service.

\subsection{Focus}
        \label{ssec:intro:focus} 
    
    The overall decision task they face is whether and how to respond with a pre-hospital critical care field team - a team that can effectively deploy the skills and equipment of an intensive care unit to a patient in the field anywhere in Wales. 
    Dispatching such a team makes possible life- and limb-saving interventions prior to any transfer to the hospital setting. 
    Ahead of any AI development work, this study aims to find out if an algorithmic assist system could provide a usable decision support tool to improve decisions and reduce variability in practice and outcomes. 
    
    At the critical care hub, the dispatching and support decisions result from a process of assessment, filtering and prioritisation. 
    Algorithmic systems must be carefully designed, validated and monitored in order to avoid unintended harms. 
    They are also subject to interaction challenges so that, for clinicians, useful insight may become lost owing to bad timing, lack of clarity or poor communication.
    So the detail of what clinicians are actually doing when they make dispatch decisions is hugely important.
    
    Combined human-algorithm decision systems are an increasing feature of the research literature in computer science. 
    But the detail of how to maximise the benefit of such combination in any given situated context is under-developed. 
    The dimensions of the combination space need to be explored more fully and the implications of these dimensions for putative design solutions need to be better understood if we are to gain maximum leverage from the involvement of algorithmic agents in decision support~\citep{wilson2025DimensionsHumanMachineCombination}. 
    Such contextual factors should also play into the architecting process of algorithmic models so that they support the kinds of interaction and convergence necessary for best combined performance~\citep{turchi2026WhenAISolves}.
        
\subsection{Prior work}
\label{ssec:prior-work}  % ssec and label added 20-Aug
    Prior clinical work has elaborated how understanding of the situated task needs to inform the design of support processes and hence algorithmic development itself~\citep{li2020DevelopingDeliveryScience}.
    Such work emphasises the need for early, iterative and situated evaluation that focuses on the human-machine combination rather than the machine alone.
    While devoting time to understanding the situated task is an expensive process for an algorithmic system developer, it is nevertheless likely to be worthwhile since it increases the likelihood that the resulting system will turn out to be useful.
    And therefore it reduces the prospect of invested time and resources going to waste.

    Previous work on the design of human-AI decision systems analyses how the dimensions of combination constitute a set of features that affect the interplay between humans, machines and decisions~\citep{wilson2025DimensionsHumanMachineCombination}. 
    It uncovers the textural richness of the space of human-machine interaction in the context of clinical decision-making. 
    At the same time, previous attempts to implement AI systems in Emergency Medical Dispatch show how decisions made in design, development and deployment can encounter the reality of human and organisational work processes and how an algorithm with demonstrably impressive performance metrics may nevertheless be unable to improve human decision-making when deployed in the real-world setting of an Emergency Medical Service~\citep{blomberg2021EffectMachineLearning}. 

\subsection{Research objectives}
\label{gemba:ssec:research-objective-work-as-done}  % UPDATE?

    The overall objective was to identify the opportunities and conditions for success for an AI assist system in the critical care hub.
    Given this long-range objective, the approach to the study of practices at the critical care hub needed to be able to uncover the fine grain of detail that existed as well as how that detail varied from shift-to-shift and even hour-to-hour.

    The critical need was to understand as fully as possible the human role in the data and decision workflow But this could not be achieved without close engagement with the situated context.
    Process improvement practitioners, safety researchers and HCI academics alike emphasise that it is essential to understand the process and the people involved \textit{in context} if you want to identify opportunities for positive change~\citep{blandford2014PatientSafetyInteractive,macleod2020ReducingWaitTime}.
    In setting out to evaluate the potential for algorithmic assistance to aid the dispatch process, we recognised this need.

    To achieve appropriate understanding, we sought to develop what David Randall, Mark Rouncefield and Peter Tolmie call `a mundane competence' in the world of the situated task~\citep{randall2021EthnographyCSCWEthnomethodology}.
    We wanted to analyse in depth what writers like Braithwaite, Thimbleby and colleagues call the `work-as-done' - not just the `work-as-imagined' or the `work-as-documented'.~\citep{braithwaite2016ResilientHealthCare,thimbleby2021FixITSee}
    As a result, we needed to place the emphasis on recontextualising decision-making `in meaningful social and material practice'~\citep{suchman2013ConsumingAnthropology}.

    Observation of the handling of calls at the service was unlikely on its own to inform design decisions since it is difficult to make sense of decision-making as an outsider.
    What decision-makers were accomplishing in the course of their actions and utterances needed to be teased out and differentiated in order to identify the types of order they were sensing, building, maintaining and repairing within their teams.

    In the language of ethnomethodology, staff who work in the situated environment of the critical care hub are not just participants. They are members of a sociotechnical assemblage in which there are observable practices and recognisable practical accomplishments.
    This paper reports the analytic practices we employed for uncovering members’ methods and identifying how they managed to reach a determination on each of the thousands of decisions they faced each shift. 
    % manage to construct intricate patterns of unfolding coordination in the course of their work.

\subsubsection{Research Objective 1 - uncovering methods}
    Our first objective was to discover the methods used by staff to accomplish their daily work.
    Because the key decision-makers also work clinical shifts, we spent time learning about different aspects of their work away from the desk.
    We studied how they referenced organisational policies and processes that were designed to structure their practice. 
    We also paid attention to how they resolved conflicts that fell out of reach of documented operating procedures.
    In particular, we noted where practitioners related how their field work was conditioned by decisions at the critical care hub.
    And how, in turn, they described their decision practice being influenced by their field experience. 
    This informed our understanding of how dispatch decisions are constituted as key moments for all members of the team.
    
\subsubsection{Research Objective 2 - identifying key decision points}
    Our second objective was to identify the pivotal decision moments that surround each dispatch decision, their inputs, influences and outflows. 
    In one sense it is obvious, even to a casual observer, that a dispatch decision is just one in a sequence of very many decisions that occur between an emergency call and a clinical intervention at the scene of an incident.
    While decisions post-dispatch were often front-of-mind for staff, we wanted to focus on any antecedent decisions as enablers and influences on the dispatch decision itself. Which steps in the process were key to understanding how a decision to dispatch or not was reached?

\subsubsection{Research Objective 3 - locating where to assist}
    Our third objective was to locate the best point in the extended process at which any form of algorithmic assist could reasonably expect to work.
    In other words, we wanted to work out where to intervene.
    And we knew this was predicated on achieving robust results against objectives 1 and 2.
    By making use of the combined research outputs and by starting with context, we anticipate some of what will become the later design processes by involving situated stakeholders in consequential deliberations.
    Deliberations on where to place development efforts and what is to be achieved by those efforts.
    This movement from context to design bench is in stark contrast to approaches that begin with technology and try to find which sociotechnical processes can be found to make use of it.
    
\subsection{Contribution}
    \label{ssec:intro:contribution} 

    In a general sense, this paper aims to encourage development efforts that are more likely to address problems in the situated context of clinical decision-making. 
    An effective decision support tool needs to be part of a socio-technical system, requiring its integration into existing social, organisational, regulatory and professional contexts. 
    The primary aim of any proposed algorithmic tool here must be to augment the abilities of the practitioners in the critical care hub so as to maximise patient benefit.
    That is, to enhance practitioners' decision-making, considering their intuition, skill, expertise and operational context.

    We begin to address a gap in the literature with this work.
    That is, by foregrounding, before any development takes place, a process to identify whether, where and how algorithmic support is likely to be effective.
    We approach the problem by first applying a process of immersive engagement (\cref{ssec:immersion-methods}) which produces a fine-grained description of the work-as-done (\cref{ssec:results-key-decision-points}).
    In particular, after a contextual grounding on the nature of triage, we are able to represent the work of hub staff as fitting into an actual workflow of distinguishable steps (\cref{sssec:the-cch-workflow,ssec:cch-workflow-in-outline,ssec:dynamic-sift-and-sort,ssec-features-of-the-call-list}) with specific contextual characteristics and challenges (\cref{ssec:challenges-of-the-workflow}) and a distinct informational profile (\cref{informational-profile-of-the-process}).
    This material addresses both Research Objective 1 (uncovering methods) and Research Objective 2 (identifying key decision points) as well as touching on Research Objective 3 (locating where to assist).

    We report a further phase of research in addition.
    In this phase of structured observation, we set out to analyse the communicative interactions engaged in by practitioners during the decision-making workflow as part of their situated work.
    The setting for this is described (\cref{ssec:observation-setting}) to provide the observation context.
    We were able to deploy ethnographic methods (\cref{ssec:observation-methods}) that capture the significance and texture of those interactions.
    In laying out the results of this phase, we first of all quantify how communication \textit{between} the two hub staff accounts for the majority of all interactions (\cref{sssec:internal-and-external-interaction}) and then show how situational awareness dominates their interactions -- especially in the form of unsolicited updates - (\cref{sssec:final-grouping-of-coded-types,sssec:patterns-detectable-in-the-observable-data,sssec:unsolicited-updates}).

    We then consider how staff interactions are affected by factors external and internal to the hub (\cref{sssec:factors-affecting-staff-interactions,sssec:variation-over-time,sssec:sources-of-variation}) and the significance of their joint curation of shared situational knowledge (\cref{sssec:shared-information}), before synthesising the accumulated findings to arrive at a more detailed picture of the three key decision steps and their interdependencies and influences (\cref{sssec:situated-action-simplifying-a-complex-process}).
    This enriched account of situated action adds depth to the material addressing Research Objectives 1 and 2.
    In particular, it will be seen that having outlined the key decision points in \cref{ssec:cch-workflow-in-outline} (\cref{fig:ecch-three-decision-steps-outline}), we have provided an account of their complex interconnections by \cref{sssec:situated-action-simplifying-a-complex-process} (\cref{fig:ecch-three-key-decision-points}).
    The depth of detail also provides the foundation for the work on our third Research Objective.
        
    We substantively address Research Objective 3 (locating where to assist) directly in its own results section (\ref{ssec:locating-where-to-assist}), drawing on the material assembled in our study of situated action.
    In this section, we consider the candidate locations for an assist to be introduced (\cref{sssec:three-candidate-steps-for-decision-support}) and provide detail on how our findings inform a view on their strengths and weaknesses (\cref{sssec:interaction-with-an-algorithm-when-time-is-tight,sssec:limits-to-algorithmic-situational-awareness,sssec:algorithmic-assistance-in-the-workflow,sssec:evaluating-decisions-in-a-dynamic-context}).
    We then briefly review the design requirements for algorithmic support, given the context (\cref{sssec:support-for-situated-decision-making}) before providing summary findings (\cref{sssec:situated-action-summary-findings}) pointing to the best candidate location for support.
    
    % and eventually scrutinise how the real-world context informs what the most effective algorithmic contribution might be (\cref{ssec:locating-where-to-assist}).

    We discuss the implication of the assembled evidence in \cref{sec:discussion} which is that one candidate step stands out as the most profitable at which to explore applying an algorithmic assist.
    And we then outline our broader methodological conclusions in \cref{sec:conclusions}.
    
    In so doing, we demonstrate the value of working from a specific situated problem to the most appropriate parameters for a potential solution.
    This emphasis on real-world contextual groundwork provides an increase in the chances of success and sustainability~\citep{oren2020ArtificialIntelligenceMedical,greenhalgh2017AdoptionNewFramework}.
    In turn, it maximises the prospect of real patient benefit.
        
    In essence, we provide an account of work done to try to understand critical features of a specific real-world setting ahead of algorithmic development and deployment.
    Operationalising responsible AI for effective workflow integration requires that developers build from context, not technology~\citep{turchi2026SixElementsSynergy}.

%% file: incl_02_methods.tex
% !TEX root = ../main.tex

\section{methods}

\subsection{Immersion methods}
\label{ssec:immersion-methods}
    Meeting our first Research Objective (uncovering methods) required work that began before, but then continued in parallel with, work on the second Research Objective (identifying key decision points) in an iterative process.
    This is because methods cannot be studied at decision instances of which the researcher is unaware.
    And not all decision points are immediately apparent to the novice observer. 
    Discovering where and how decision moments are encountered and addressed requires the researcher to have access to practitioners' methods of sense-making, communication and application.
    In turn, identifying what kind of decisions are made, and where, enables further exploration of practitioners' methods, so each iteration advances both objectives.
    
    This was accomplished during an extensive period of exposure to the practices of critical care hub staff, and to the ways their activities fit into the overall setting of the Emergency Medical Retrieval and Transfer Service.  
    The researcher attended work meetings and training days to hear active discussions of operational issues and how they were being addressed month-by-month.
    Time with practitioners was structured to enable the researcher to encounter a diversity of uncontrived settings and experience an immersion in the context  of practice.
    This included arms-length observation to enable analysis of dialogue and interaction both within and outside of the immediate work environment.
    Our interest in practitioners' own methods meant the researcher was keen to discover now they coordinated action, how they accomplished communicative repair and made sense of situations.
    
    This work involved focusing on the study of what ethnomethodology calls accountable procedure.
    Accountable in this context means self-explanatory, visible, describable and intelligible within the situated context -- giving an account of itself.
    For this to be possible, it required a familiarity with the multiple tasks undertaken at the critical care hub (hereafter referred to as simply `the hub').
    Superficial knowledge of how a dispatch decision is executed could easily miss the significance of associated tasks and challenges as well as the numerous influences on the decision itself.
    But it would also mean that many observed actions could be accountable, in this sense, only within the team -- at the same time being unintelligible to the researcher.

    Immersion involved getting to know the details of team language as well as tools and resources employed. 
    Operating procedures and training materials provided useful reference points for recognising elements of practice and how they fitted into the processes and workflow encountered at the hub.
    But they were not a primary source of evidence.
    Alongside documentation, we were able to access high-fidelity anonymised call lists and training environments in order to understand the nature of the information being processed during hub work on call lists.
    Each of these sources was enriched with detailed discussions with the hub manager, the educational and research leads and practitioners - both in and out of the workplace.
    
    Discussions, documentation and data were informative but secondary.
    To understand the `work as done', we needed to ensure time was spent observing staff over several shifts at the critical care hub itself.
    We obtained ethical approval for unobtrusive observation using pen-and-paper recording.
    Our research on methods and key decision points was conducted over iterative phases, but took essentially two forms, unstructured immersion and structured observation.
    Each was supplemented by extensive informal discussion with staff away from their work pressures so that sense could be made of what had been observed.
    The plan for the later, structured observation (\cref{ssec:observation-methods}) was formulated in the course of analysing and writing up results from immersion and from analysis of key decision points.
    And the findings, including on methods and processes, are incorporated into results detailing the decision points (\cref{ssec:results-key-decision-points}).
    
    This foundational phase of immersion enabled the researcher to pick up key elements of technical and operational processes, how they were navigated in practice and how they were blended in different ways on each shift.
    An opportunity was presented to witness how, alongside their interactions with systems, pairs of practitioners used fragments of information passed between them as part of a coordinated and mutually supportive maintenance of shared situational awareness.
    Researcher immersion in the work environment amounted to over 50 hours, including five shift handovers -- during which a concise but explicit commentary is always shared between staff -- and this meant that uncertainties in interpretation were clarified.
    This was in addition to a further involvement in approximately 50 hours of discussion in both formal and informal meeting settings.

    The results of this work provided the detailed contextual grounding for addressing both Research Objective 1 (uncovering methods) and Research Objective 2 (identifying key decision points).
    The findings are reported below (\cref{ssec:results-key-decision-points}).

    % In what follows, we describe the work of the hub staff at some length while attending to the fine detail of decision-making processes.

    % Recalling the two components of our objective in studying the situated work of hub staff (\cref{gemba:ssec:research-objective-work-as-done}), we have dealt with the first - identifying the key decision points in the workflow - and now turn to the second.

\subsection{Observation setting}
    % \label{gemba:sec-approaching-observation-at-the-hub}  % check for refs that are now broken
    \label{ssec:observation-setting}

    For structured observation, detailed observational records were made over three day shifts and two night shifts with manual recording (using a derived coding system) of 22 operational hours. 
    These observations provided the basis for detailed descriptions of the task and workflow interactions engaged in by hub staff as described in \cref{sssec:situated-action-simplifying-a-complex-process}.

    The purpose of a new phase of structured observation at the hub was to add both qualitative and quantitative texture to the material so far established under both Research Objective 1 and Research Objective 2. 
    That is, having identified the key decision points, to address in detail an element of the objective of uncovering methods -- how staff reflexively negotiate the key decision steps.
    How do they accomplish their determinations at these points, marshalling inputs, influences and outflows? 
    And, in doing so, how do staff interact with each other, with the dispatch system and with other information flows?

% \subsection{coding}
    The researcher's significant experience of working with clinical teams and conducting ethnographic studies facilitated a rapid and efficient process of coding development.
    An iterative development of the paper-based recording technique was accomplished by means of an initial observation-recording session lasting five hours during which coding approaches were derived, applied, evaluated and refined.

\begin{figure}
        \centering
        \includegraphics[width=8cm]{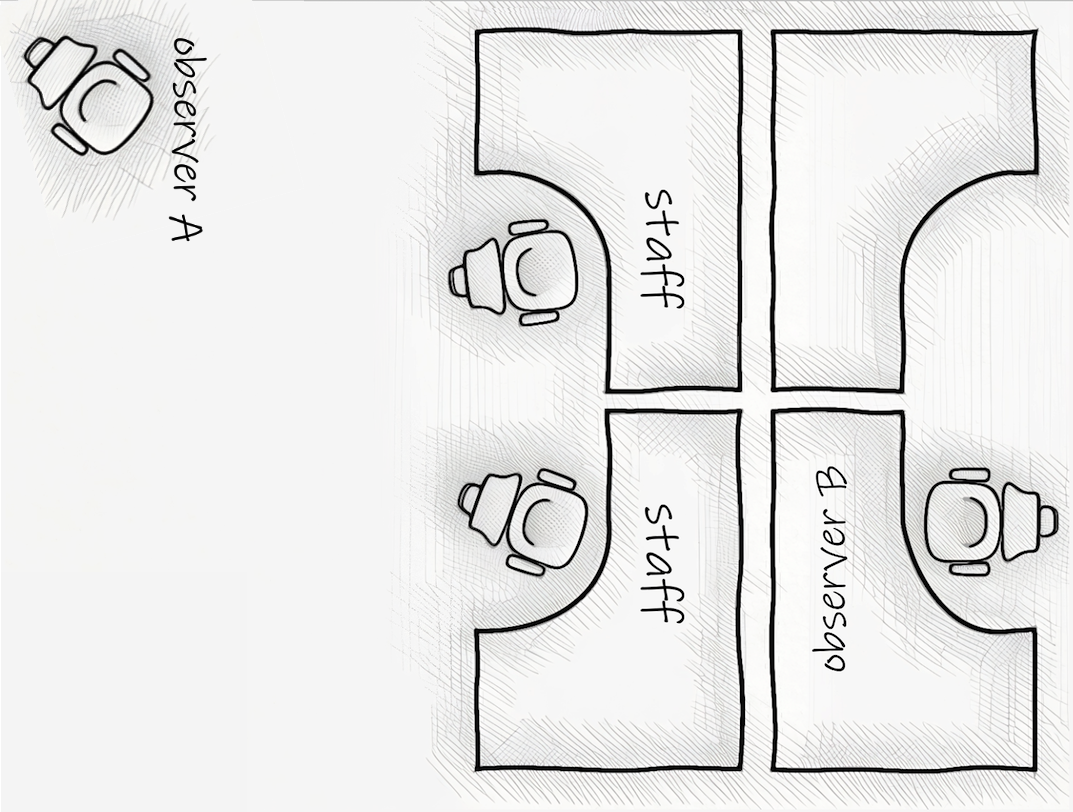}
        \caption{Plan of the observer's two different orientations to the hub staff -- labelled as observer A and B}
        \label{fig:observation-plan-sketch-03a}
    \end{figure}

\begin{figure*}
        \centering
        \includegraphics[width=\linewidth]{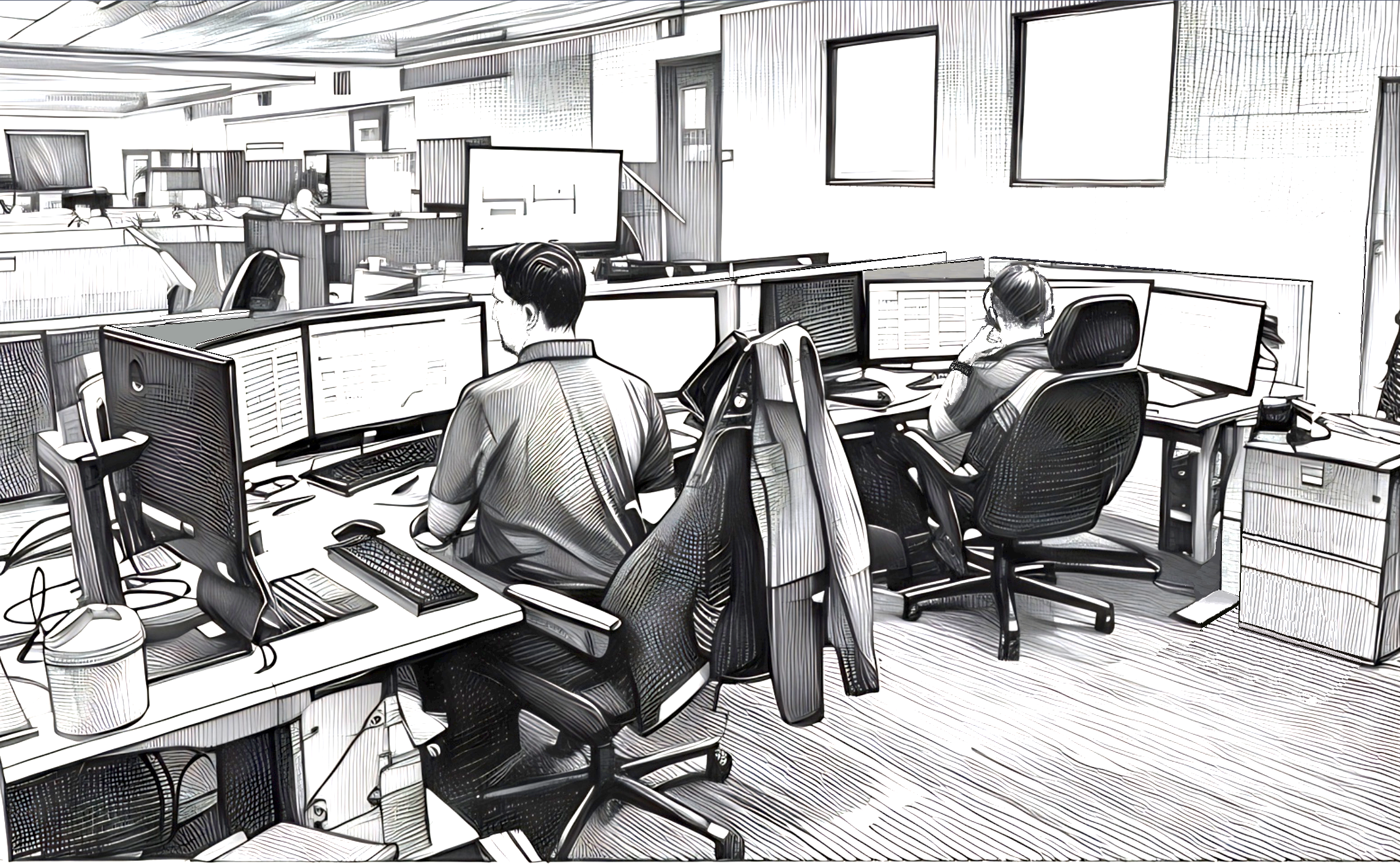}
        \caption{Sketch of the observer's view from plan position A.}
        \label{fig:sketch-03}
    \end{figure*}

\begin{figure*}
        \centering
        \includegraphics[width=\linewidth]{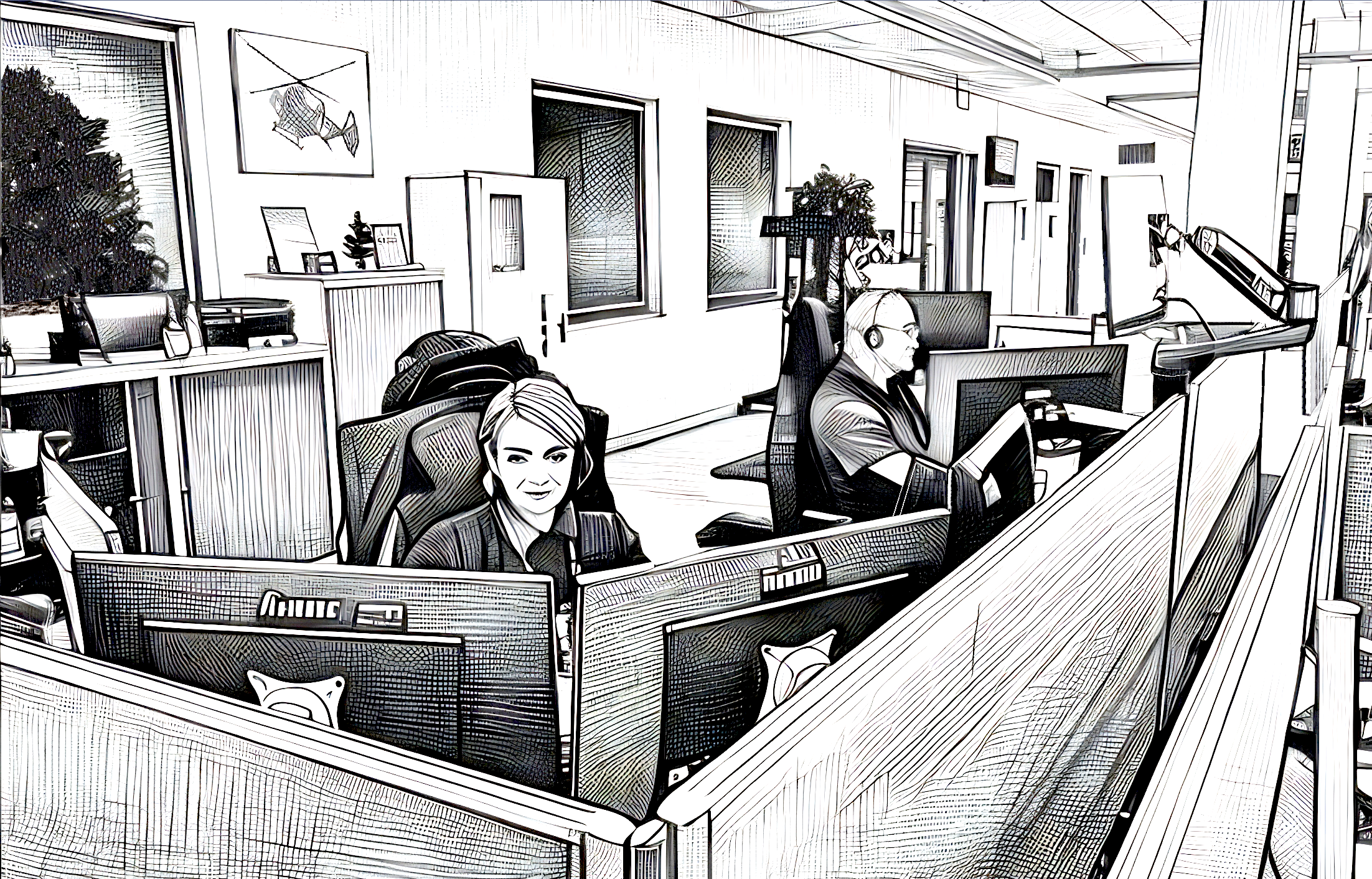}
        \caption{Sketch of the observer's view from plan position B.}
        \label{fig:sketch-02}
    \end{figure*}

%\subsection{Positioning the observer}
    To comply with ethical approvals, observation was carried out from positions where screen information was not readable.
    Having established what would be a safe distance from the front of the screens (observer A in plan \cref{fig:observation-plan-sketch-03a}, providing the view as seen in \cref {fig:sketch-03}), some of the time was spent positioned so that the use of paper-based notes by the two staff could be observed.
    However, most of the observation was conducted from a position facing the two staff (observer B in \cref{fig:observation-plan-sketch-03a}, so that the two staff are seen as in \cref{fig:sketch-02}).
    While this latter position obscured the use of paper-based notes by hub staff, it made it possible to observe more closely where their attention was focused and, in particular, when each was attempting eye contact or looking across to see where their colleague’s attention was directed.
    
    % \description{       
    % A sketch plan showing two adjacent staff desks and two different observer positions. The position labelled `observer A' is some distance behind one of the staff positions. This gives a view of the two working areas and screens, while being too far away to see any detail. The position labelled `observer B' uses an immediately facing desk so that interactions between the two staff are visible to the observer and the observer, in turn, is in their upper line of sight. 
    % }
    % \description{       
    % A sketch from behind two hub staff at adjacent desks in a large open-plan office. There are 12 information screens arrayed in front of the two staff (seven in front of one, five in front of the other). These include laptops and ipads that are in use. One of the staff team is wearing a telephony headset. The two are looking at their screens, not at each other.
    % }
    % \description{       
    % A sketch from in front of two hub staff. Each is wearing a telephony headset. The two are looking at their screens, not at each other. The same array of screens is visible. But the observer is closer to the staff. Their faces are clearly visible so that non-verbal signals would be more readily detected.
    % }
    Observation under these conditions does not consider the specific human-computer interface of the screen with its list of calls and the mouse with which an individual call can be opened and explored.
    Instead, it foregrounds the attentional and interactional patterns displayed by hub staff as they navigate the flow of incoming information.
    This foregrounding is important as such patterns always provide the cognitive backcloth for each interaction with the on-screen list.
    Staff on the desk are not able to concentrate on the call list to the exclusion of all else.
    They don’t work in cloistered silence with a definitive list to peruse for candidate cases.
    Their reality is one of continual dialogue with operational colleagues, with people at incident scenes, with each other and with fellow teams in the call centre beyond the hub itself.
    Between all these dialogues, they must scrutinise a call list that is long and constantly being updated.
    For this reason, our focus here is primarily on speech dialogues - albeit with an awareness that people's actions beyond words are also part of their workflow.
    And we seek to discover what these speech practices together reveal about what they are doing and thinking.

%\subsection{Attention, task, dynamics and interaction}
    We were especially interested in how attention is given to the different tasks performed by hub staff and how this is accomplished between them.
    What are the dynamics and processes of the two staff working together?
    To what extent are they operating in parallel and how much they are doing joint work?
    We spent time considering how their interactions were conducted in different interaction types – interactions for specific purposes with each other, with the call list information, with the operational teams outside the room and with other colleagues who were inside the room but not themselves in the hub function (ie, ambulance service staff).
    The relevance of this exploration for the design of any support technology is that the critical information flows need to be understood.
    The attentional switching regularly performed by hub staff and their own indication of the information they're seeking or supplying in the course of accomplishing a task tell us a great deal about what sort of assistance they might best be able to exploit to particular advantage.
    For a given decision, the more timely, complete, consistent and reliable any assistance is, the more it will be useful to a human decision-maker.
    Furthermore, if attention itself and attention-decisions together make up a significant part of the real workflow, then whether these are primarily joint or individual enterprises of the two human operators is of great consequence.
    
    Our previous examination of the work-as-done was informed by a level of familiarity with both the work context and the character of information flows.
    Our focus for this new phase of observation was the degree and patterning of collaboration between the two staff and how this was negotiated in the course of a shift.
    We were able to employ mixed methods that leveraged formal ethnographic techniques to uncover significant insights into the ways in which the hub team accomplish and sustain their work and how this varies over the course of different shifts.

\subsection{Observation methods}
    \label{ssec:observation-methods}
    
    The researcher used the first of the five hours of observation to pursue an open coding approach using the constant comparative method.

%\subsection{Coding initiation}
    Glaser describes the constant comparative method as a means to address the paradox of coding qualitative material as otherwise being a disordered sequence of theory-testing and theory-generation~\citep{glaser1965ConstantComparativeMethod}.
    By interleaving the two processes in an iterative, continual (or constant) comparison of the current code instance with many others of nominally the same or a contrasting type, it becomes possible to give space to idea-generation while at the same time exposing the ideas to continual testing.
    This first hour yielded a coding system that centred on how the two-person team functioned.
    
    Having begun with codes that foregrounded the different modes of human-human interaction (speech via telephone, radio or face-to-face) alongside human-computer interaction, our coding evolved in light of the dominance of the different interactions directly between the two staff.
    We ended the first hour with a system that primarily coded the different forms of the latter and separately coded the interactions they conducted either individually or together with the world beyond the hub itself.

    \input{tbl_obs_interaction_ix_construct}

\begin{figure*}
        \centering
    	\subfloat[Generic]{
    		\includegraphics[width=0.478\textwidth]{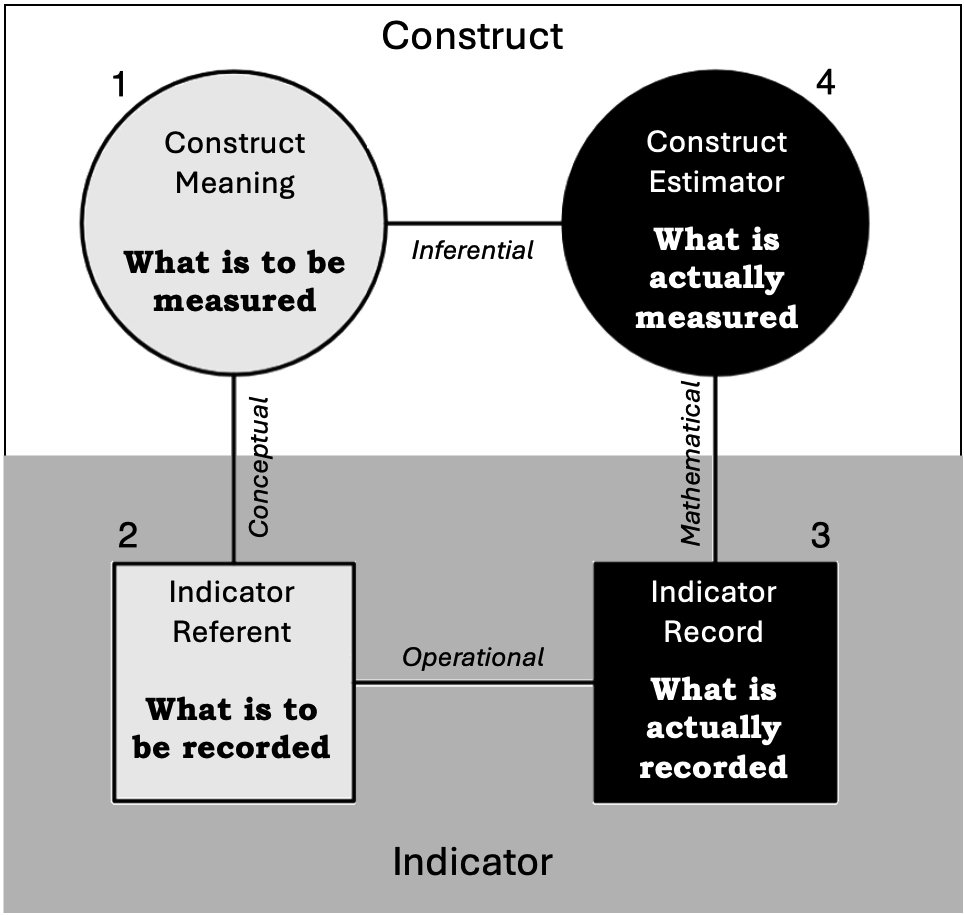}\label{fig:constr-indic-link-a}
    	}\hfill % Spacing between sub-figures displayed next to each other.
    	\subfloat[Context-specific]{
    		\includegraphics[width=0.478\textwidth]{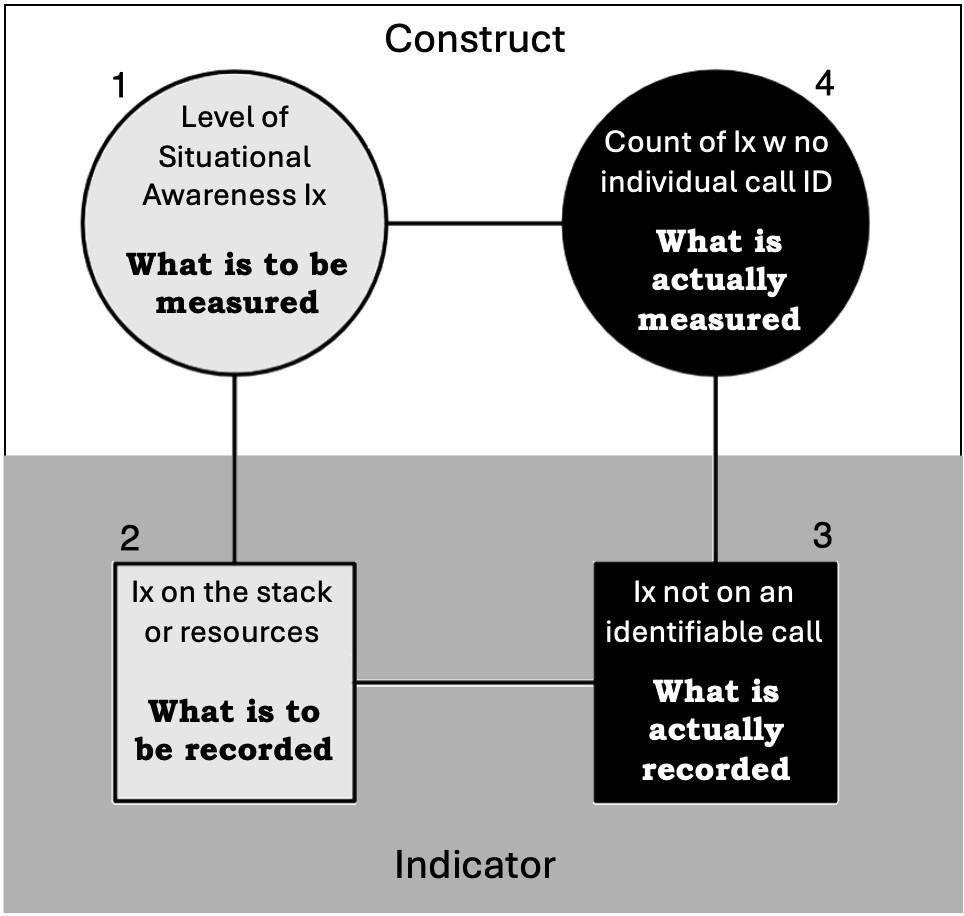}\label{fig:constr-indic-link-b}
    	}\\ % New line before caption.
        
        \caption[Zwanenburg's construct-indicator linkage model]{Zwanenburg's construct-indicator linkage model~\citep{zwanenburg2015HowTieConstruct}.}
        \label{fig:construct-indicator-linkage}

    \end{figure*}

    A review of this work evaluated the emerging construct-indicator relationships~\citep{zwanenburg2015HowTieConstruct} to confirm the validity of assumptions made in the coding process.
    \Cref{tbl:obs-interactions-on-the-list} lists Zwanenburg's four generic quantities against four contextually-specific descriptors that help to emphasise the attention we need to pay to the detail of measurement recording.
    If we begin with what we want to measure and consider how we can possibly record what is witnessed, we see they are necessarily distinct entities (and potentially distinct quantities).
    The actuality of recording makes a further alteration.
    And what that constitutes as an actual measure can then be compared with the original measurement intent.
    The attempt to measure thus undergoes three transformations (\cref{fig:constr-indic-link-a}).
    A conceptual linkage links the idealised measure (1) to its referent in the conceptual record (2).
    An operational linkage transforms this into the concrete actual record (3).
    And finally, a mathematical linkage transforms what is recorded into a resultant actual measure (4).
    What has been accomplished is a real measure that is an estimator for the target meaning of the contstruct. 
    And we confirm that it is sufficient as an estimator by scrutinising the linkages.
    \Cref{fig:constr-indic-link-b} shows how Zwanenburg’s three measurement relationships (conceptual, operational, mathematical) combine to produce the resultant (inferential) relationship between the target meaning of a specific construct and the estimator of that construct.
    
    We show in \cref{fig:constr-indic-link-b} how this process applies to our attempt to measure the occurrence level of interactions (abbreviated as `Ix') concerning situational awareness.
    The referent of the indicator in this example is an interaction between the two hub staff on either the call stack itself, cases in general or on the resources at their disposal.
    This being the level of specificity at which situational awareness exists – ie, everything within the work of the hub that is above the internal detail of an individual, identified case.
    The researcher’s record of an indicator is a note of an interaction in which there is neither implicit nor explicit identification of a specific case – eg, neither a (previously specified) cardiac arrest, nor a (previously specified) case in a given location, nor the case with incident number specified as ending 4321 and so on.
    Our estimator of the target (the level of situational awareness interactions) is then the count of codes recording such events.
    
    We can see that the scope for errors, omissions and misidentifications exists in all three relationships.
    The point of this construct-indicator review was to attempt to minimise these issues by checking for the occurrence of events where the conceptual relationship was weak or difficult to justify.
    At the same time, it was an opportunity to confirm that what was actually noted on paper would be a consistent and readable record of the indicator.
    
    In the case of situational awareness interactions, the conceptual relationship was weakened to a degree if the two staff were each listening in on the same call and began an interaction in which their shared awareness of the case identifier was already accomplished, but at the same time obscured from the observer.
    However, during this review, the researcher was able to establish that it was rare for the two staff to be listening in on the same call without that fact being clarified or confirmed between them through some utterance.
    Indeed, because each was using their own headset, they would invariably make a point of asking whether the other was listening in on the same call – as identified by some feature combination specific to the case.

%\subsection{Coding refinement}
    A further hour of observation using this refined coding was utilised to develop an axial coding where both super-classes and sub-classes of existing codes were elaborated.
    During this process, the different categories of solicited and unsolicited utterances (that feature later) became refined and embedded.
    This is an observer-identified distinction that arose as a result of paying attention to the (fragmented) flow of information passing between the two hub staff.
    
    A final hour of observation was then used to check linkages and refine the coding system~\citep{hammersley2010EthnographyPrinciplesPractice}.
    This involved searching for contra-indicated cases, comparing across different interaction types, checking assumptions and cross-referencing earlier discussion notes.
    
    The hub manager was available to answer questions for most of this initial session, which meant that uncertainties could be ironed out with minimal delay.
    And this meant that time was also available to reflect on the role of the researcher in selecting the data, the codes and the likely signals they would capture.
    In particular, attention was given to the ways in which the interesting dynamic between the specific people working the shift together as the hub team might appear to alter the type and quantity of language used in their interactions.
    Language is a technology, too~\citep{wegerif2019BuberEducationalTechnology}.
    One that is socially constructed and individually implemented.
    So different pairs of people might produce different patterns.
    Would terse exchanges lead to crisp and concise notes and unambiguous codes?
    Would verbose narrative descriptions lead to meandering notes and an uncertainty in coding?
    In the end, it became clear that the active refinement of coding categories would provide the best assurance of recording quality.
    
    Having satisfied the requirement of iteratively refining and evaluating the coding system, the researcher conducted a further 17 hours of detailed observation of critical care hub staff at work.
    Analysis of the coded observation notes reveals that interactions, while not uniformly spread over the hours of a shift, occur at a rate of one every 2 minutes on average.
    This might seem to point towards a challenge for maintaining concentration on screen-based information over the twelve hours of a shift since such attention is regularly interrupted in these conditions.
    However, the efficiency with which hub staff are observed to anticipate their colleague's focus means that they are each frequently ready to interact on the same issue at the same moment.
    This itself must inform the design of any assist tool.
    Attention is rarely on a single task for long unless that task is a call reasonably close to a dispatch decision.

    Together with collaborative design discussions, the detailed familiarity achieved here informed the approach to experimental work which is the subject of a forthcoming paper.
    We conclude the methods section with a note that the presentation of results differs from the sequence of their acquisition.
    Developing the identification of key decision points is the first focus, while the complex connections between them are delineated as a description of situated action, once interactions have been analysed.
    In reality of course, the research processes are neither linear nor neatly isolated, as we note next.

%% file: tbl_obs_interaction_ix_construct.tex
% !TEX root = ../outline.tex
\begin{table*}
	\centering
	\small
	\caption[Construct-indicators for interactions on list situational awareness]{Construct-indicator tabulation for interactions on situational awareness on the call list (ie, on the `stack')}
	% \begin{tabular}{p{0.45\linewidth}|p{0.45\linewidth}}
	\begin{tabular}{l|l}
		\toprule

Generic quantity & Contextually-specific quantity\\

		\midrule
        
What is to be measured & Level of situational awareness IX (interactions)\\
What is to be recorded & Ix (interactions) on the stack or resources\\
What is actually recorded & Ix (interactions) not on an identifiable call\\
What is actually measured & Count of Ix (interactions) w no individual call ID\\

		\bottomrule
	\end{tabular}
	\label{tbl:obs-interactions-on-the-list}

\end{table*}

%% file: incl_03_results.tex
% !TEX root = ../main.tex

\section{Results}
\label{results}

\subsection{Results 1 - key decision points}
    \label{ssec:results-key-decision-points}

While our Research Objectives appear to have a natural order (uncover methods, identify key decision points, locate where to assist), the reality is that the work proceeded through a series of iterations in which each of the three objectives received focus in turn.
Evidence accumulated in layers in relation to all three in a process of imbrication.
We therefore present our results as an account of the key decision points.
And in the course of detailing the workflow we report on our findings in relation to practitioners' methods and opportunities for algorithmic assist.

    We thus begin here with the nature and context of the decisions to be supported.
    And in our case, the nature of critical care dispatch decisions is that they are substantively executed only after a significant number of prior steps - some of which are information-gathering, synthesising and analysing steps, but many of which are distinct, preparatory decisions that select candidate cases from of the bulk of calls coming in to the team.
    The fact that these steps can happen very rapidly does not alter the fact that they condition any substantive decision on dispatch of critical care resources.
    This preparatory selection process is known as triage and understanding how it characterises the work is critical.
    So we will explore it in some detail before reporting on our specific context.

\subsubsection{Contextual grounding - the role of triage}  % poss move to incorporate in section title?

    % \label{gemba:sec:intro-method}
    \label{gemba:sec:intro-method-triage}

    For healthcare systems worldwide, clinical triage is a decision process that aims to prioritise care while optimising resource allocation. 
    The most consequential triage processes are those that dictate action in medical emergencies. 
    Decision-making in this context is time-critical and complex, using incident history and clinical signs to sift and sort cases into identifiable and meaningful groups.
    Specifically, triage is a preparatory decision process - one decision made in order to facilitate, improve or expedite another subsequent and substantive decision.
    
    Emergency medical triage takes place in a number of different contexts, each with its own specific challenges. 
    In each case, the burden on human decision-makers can be immense due to the pressures of volume and complexity and the rapidity of case-switching.
    While in the hospital Emergency Department, direct observation enables immediate triage, continual review and re-triage if necessary, in the case of ambulance dispatch, the initial information and consequent action may be separated significantly from any update. 
    For staff in the emergency department, there is the simultaneous pressure of being co-located with a sometimes large collection of ill and injured people needing both care and continual monitoring. 
    For ambulance dispatchers, there is a different burden of not being able to provide direct care from their remote location. 
    In either case, inevitably, there is additional psychological pressure if there are too many cases for the available resources.
 
    Critical care triage is conventionally carried out on arrival at a hospital site, some time after an ambulance dispatch process.
    In our specific context, \textit{pre-hospital} critical care dispatch, decisions are made on tasking specialist teams to provide high-level care interventions \textit{at the scene of an incident} - interventions of the type normally carried out only in a hospital Emergency Department itself. 
    Dispatch decisions for this kind of service are a special subset of clinical triage processes. 
    They are unlike those in a typical Emergency Department because pre-hospital critical care teams are dispatched to the patient, so several steps are executed while the clinician is remote from the patient. 
    But they are also unlike a typical ambulance dispatch response because they are highly selective.
    There is a much greater emphasis on what significant clinical interventions can be carried out on scene, and what difference can be made to clinical outcomes by means of pre-hospital critical care. 
    Where pre-hospital critical care is not indicated, the case is still attended. But it is dealt with by the wider ambulance service through its normal dispatch process.
    
    As mentioned in the introduction (\cref{ssec:intro:context_em_med_dispatch}), in Wales, UK, pre-hospital critical care is provided from four helicopter and rapid response bases by the \textit{Emergency Medical Retrieval \& Transfer Service, Cymru}. 
    The specialist dispatch process is coordinated at a single critical care hub where two people are responsible for monitoring all emergency medical calls throughout the country.
    Decisions on dispatch of pre-hospital critical care are made by a clinician in this service.
    The clinicians involved also spend most of their shifts each week serving within clinical field teams. 
    So their decision-making practice on dispatch is uniquely well-informed and hence highly advanced.    
    This makes the Welsh context particularly useful as a place to explore how algorithmic assistance might support the dispatch process for the wide range of incidents they cover.
    
\subsubsection{The critical care hub workflow}
    \label{sssec:the-cch-workflow}
    
    Our work with practitioners revealed that the system-level decision to dispatch or not is, in reality, accomplished through a multitude of discrete steps - each being a decision in itself - many groups of which are often iterated over and over as new information or new circumstances prompt a re-appraisal of an earlier, provisional determination.
    At the core of this frequently-recurrent, continually-evolving and multi-threaded decision process lie at least three key decision-steps (\cref{fig:ecch-three-decision-steps-outline}).
    The first two are triage steps that enable a third and final decision on dispatch to be made on a tiny fraction of the total call list. 
    Each key triage step acts to reduce the volume of calls that have to be processed at the next step.
    
    While revealing these key decision steps and their place and significance in the overall workflow allows us to document \textit{what} is done, we also aimed to describe \textit{how} this is accomplished in order to understand what the human decision-makers are bringing to, and taking from, each decision-step - and to anticipate the impact of potential change on the humans involved.

    One of most striking things an observer notices on seeing the staff work at the hub is that the pair of them each has many systems to interact with.
    There are at least four full-sized screens plus a tablet device each.
    And each uses both a telephone and radio as well as talking to each other and people within the wider control room. 
    These are people who already interact with multiple computerised systems and are constantly accessing and processing huge volumes of information of different kinds.
    A sketch of the work environment is included above (\cref{fig:sketch-03}).
    
\subsubsection{The workflow in outline}
    \label{ssec:cch-workflow-in-outline}
    
\begin{figure*}
        \centering
        \includegraphics[width=\linewidth]{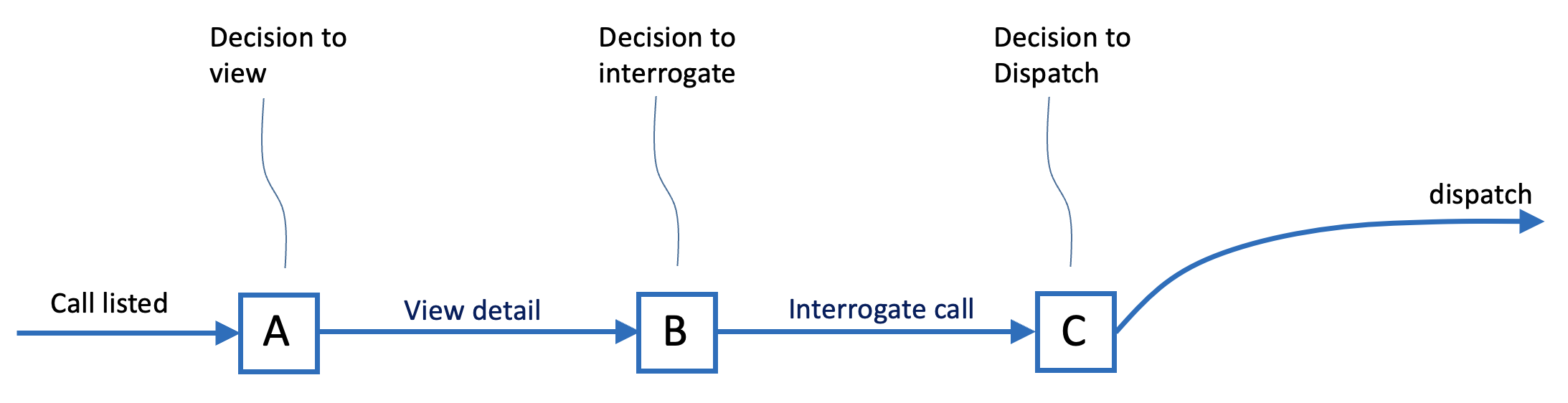}

        \caption{The three key decision-steps in outline}
        \label{fig:ecch-three-decision-steps-outline}
    \end{figure*}

    The hub staff must review the list of incoming emergency calls and decide whether a given call is such that it justifies action from their specialist clinical colleagues.
    Importantly, the main ambulance service dispatch process will review and take action on every single case. 
    The hub's distinct task is to find the \textit{subset} of calls that are likely to benefit from \textit{pre-hospital} critical care - a very small proportion of the whole list.
    A precondition for action by a critical care team is further exploration of the case since tasking a field team cannot be carried out on the basis of the listed headline alone.
    The first step in this further exploration is a \textit{decision to view} the call - that is, the call is opened in order to view it (\cref{fig:ecch-three-decision-steps-outline} A).
    This is achieved by a mouse-click on a single list item, which causes an expanded window with further detail on the individual call to be revealed on a separate screen.
    This expanded view can, in turn, give access to a further level of detail if required.
    But the information most relevant to an immediate decision is that accumulated during the emergency call itself, which is visible at the level first opened.
    
    This call dialogue information is recorded by a combination of automated logging and human-entered textual description into a record called the `sequence of events'.
    The person entering this description is an ambulance service call-taker sitting elsewhere.
    From this, staff are often able to glean information that was provided during the earlier course of the call.
    However, it is rare that staff find it justifiable to spend time scrolling through the sequence of events unless there is a specific target for their search.
    
    Following a call having been opened and viewed, a second decision needs to be made on whether the situation justifies further exploration.
    This is a \textit{decision to interrogate} - or seek information directly from scene (\cref{fig:ecch-three-decision-steps-outline} B).
    If this is justified, then it can often be accomplished by listening into the call in progress.
    This step increases both the quantity and pertinence of information upon which a critical care dispatch decision could be made.
    Hub staff do not join the call themselves.
    But they can insert questions or advice into the system that are then visible to the call taker.
    In this way, they may influence the dialogue without interrupting the normal audio interaction.
    Another approach is that they may request that the call is passed over to them once the call-taker has concluded, so they can ask further questions not already covered by the call-taker's protocol.
    Through these means, they may influence the care and advice given to the patient by providing advanced level direction and decision support to those on scene even if a critical care dispatch decision is never seriously considered.
    
    In addition to the sequence of events log in which questions and answers are noted, there is an audio log of each call.
    But the time this takes to review means it is only suitable for \textit{post hoc} auditing and case investigation.
    It is rare that staff are able to abandon the current flow of information in favour of a search elsewhere in the sequence of events.
    So the ongoing, real-time call audio is invariably the primary source of information at this point.
    
    If a call of interest is no longer in progress, then staff may find it possible to call back to the scene of an incident.
    This is most frequently done by opening dialogue with an ambulance crew already on scene, or in anticipation of one imminently arriving.
    It may be that some other first responder is accessible through the integrated communications system.
    Or a bystander or relative may be contactable at scene.

    % \description{       
    %     \item An overall flow is shown from left to right passing through three nodes, A, B \& C. Flow of cases into the first node (A) is labelled as `call listed'. Node A is also labelled as `Decision to view'. Flow of cases from the first to second nodes (A to B) is labelled as `View detail'. Node B is also labelled as `Decision to interrogate'. Flow of cases from the second to third node (B to C) is labelled as `Interrogate call'. Node C is also labelled as `Decision to dispatch'. Outflow from node C is labelled as `dispatch'.
    % }    
        
    This process of interrogating information available directly on scene, leads to a third decision point - \textit{a decision to dispatch} (\cref{fig:ecch-three-decision-steps-outline} C).
    Does this call justify mobilising critical care resources out into the field to improve the likelihood of a positive outcome?
    In some cases the decision on mobilising critical care resources is made without interrogating the information direct from scene.
    While there are rare cases of this `immediate dispatch' type being warranted, where the interrogation step is omitted for speed, most dispatches are ‘interrogated dispatches’, where activation follows on from a more or less extensive consideration of the details of the case.

    For completeness, there is a minor post-script to this workflow.
    The decision on \textit{whether} to dispatch a field team, is made by a clinician at the hub as the centre of operations.
    On the other hand, \textit{how} that dispatch is accomplished, whether the field team travel by air or by road, is a decision made by their team lead in conjunction with the pilot, whose decision on flying conditions always has priority.
    Moreover, it is frequently the case that any subsequent transfer to hospital is more safely completed with the wider ambulance service (ie, by road) rather than by aircraft.
    The critical value of the helicopter fleet is in getting the specialist pre-hospital team to the scene itself, where they can have the greatest beneficial impact.

\subsubsection{Dynamic sift and sort}
    \label{ssec:dynamic-sift-and-sort}
    
    The combination of sift and sort is familiar in medical triage. 
    Sifting is a classification task, placing cases into distinct groups according to the most suitable care pathway.
    Sorting is a prioritisation or ordering task akin to the function of mathematical regression where various features determine a monotonic dependent variable indicating priority order.
    The distinctive features of our context are the instability of the results alongside the the high selectivity involved.
    Emergency medical triage decision outcomes can remain provisional and subject to continual change throughout much of the process - up to the point where resources are committed or an intervention begins.
    Calls tend to evolve as the clinical condition of a patient changes, or their signs and symptoms become more clear.
    And each call's relationship to other calls on the list may change with each new addition to the list itself.
    The experience of hub staff may tell them that a given call should be re-visited in case an update would alter their initial decision on critical care support.
    Staff will frequently ask that on-scene ambulance crew or third parties should call back if a condition changes.
    But their experience and training often leads them to be pro-active in calling back to scene when a case looks like it might become one suited to critical care support.
    
    This need to await an update may be recognised very early on, as a call is coming in.
    Or it might be seen because of the need for some on-scene clinical update – for example from a regular ambulance crew, once they have been able to assess the situation first-hand.
    The early situation at-scene is frequently subject to some confusion and while the initial description of signs, symptoms or mechanism of injury might be sufficient to begin alerting an on-duty team to the likely prospect of an activation, the hub staff know that it is important to try and get confirmation of certain details if that’s possible before making the dispatch decision.
    This caution is essential to balance the risk of dispatching too soon and finding the resources are unavailable for another call when they could have been more beneficially assigned there.

    To emphasise what the overall challenge is, then, we have a relatively straightforward sift and sort task that is rendered complex by the fact that it has to be accomplished over three distinct steps with three distinct sets of input data streams (text, audio/audio-visual, multi-modal), and with the additional challenges of both dynamic input data and dynamic response capability.
    The whole problem space appears well-suited to algorithmic modelling because of this complexity and dynamism.
    Whether this appearance is sustained on closer examination is a question we want to address.

\subsubsection{Features of the call list}
    \label{ssec-features-of-the-call-list}
    
    For this complex process, the incoming call list is the starting point of all activity.
    Its design and operation is therefore worth describing.
    As with many of the communications and coordination systems available at the hub, the call list is part of an interdependent set of technologies commissioned and maintained by the wider ambulance service.
    This means that developments, upgrades or interoperability improvements are, for all intents and purposes, only possible at a pace and within budget determined by the needs of the wider ambulance service itself.
    The exception to this being that the system supplier provides an application programming interface (API) to a web-based version of the list that raises the potential of adding a limited level of algorithmically-sourced content at the presentation layer without significant difficulty.
    
    So much for the current technology and the sociotechnical limitations on introducing change.
    Its design and operation still need to be described - we need to understand its informational features and how it fits into the workflow.
    The rest of this section addresses these questions with insight gained from studying the work-as-done.
    
    Each entry on the list is created by a call-taker elsewhere in the contact centre and this occurs whenever they pick up an incoming emergency call (999 in the UK).
    Each entry appears as a single line of information for an individual call as shown in \cref{fig:cad-stack-alerts}.

    For a call to come to the attention of the desk staff it must first appear on the call list.
    While it's possible that a given call is brought to someone's attention by a colleague, that colleague will have first seen the call on the call list.    
    The three-step decision process begins here.
    And because that process can be initiated many times every few minutes, interactions here can have a significant impact on workflow and task efficiency.
    
    At this point we see a feature of work on the hub desk that is distinct from the multitude of workstations elsewhere in the clinical contact centre - where dispatchers of the ambulance service do what looks like a similar job in dispatching resources to calls on the list.
    Whereas an ambulance service dispatcher will have a specific geographical area (a part of Wales) which they serve, the two staff of the critical care hub cover the entire country of over three million people.
    As a specialist service they need to rely on a process of rapid sifting over the entire list to identify cases that could benefit from their specialist support.
    It’s not practical for them to switch between regional views that might contain a handful of calls each.
    So their view is of a considerably longer and more rapidly updated list.

    As a consequence of their list being longer, there are usability challenges.
    In a sufficiently short list, the cognitive burden of keeping or finding your place is low.
    And this is only slightly raised if a proportion of calls is updated and moves to a new position in the list as a result.
    As soon as the list extends beyond 15 or 20 items, the process of keeping track as one consults other screens or deals with interruptions and returns to the list causes a significant increase in cognitive burden.
    When the list reaches over 50 items, the burden of switching attention away and back again becomes significantly interruptive as the visual cues that help the user to re-orient become relatively fine-grained and harder to distinguish.
    Add to this the proportion of calls that update and change position and the higher absolute count of distractor effects can take a significant toll.
    
    In the course of a day the list may have a varying number of calls.
    As mentioned, calls come onto the list as a result of a call-taker accepting an incoming call.
    Calls are removed from the list once the attending crew or the call-taker determines that no further assistance is needed.
    So the ‘stack’ grows if calls are added more rapidly than they are dealt with and its size therefore depends on the volume and rate of incoming calls as well as the availability of ambulance crews to attend and resolve.
    At any given time, if it's quiet, there may be 100 calls visible on the all-Wales list.
    If it's a busy period, there could easily be 250 calls present at once, fewer than 50 of which can fit into the visible height of a screen.

\subsubsection{Challenges of the workflow}
\label{ssec:challenges-of-the-workflow}  

    The existing call listing system, being designed for the wider ambulance service, is not well-suited to the work of the critical care hub. Interviews with both managers and hub staff revealed that previous attempts had been made to modify the list presentation so that it would better suit the hub staff. 
    These included the introduction of a functionality in the list that allowed call-takers to flag up calls they thought to be of interest to the hub team.
    Unfortunately, the difficulty that call-takers had in detecting cases that could benefit from pre-hospital critical care made it too onerous.
    The flagged list contained many false positives and missed many cases as false negatives and so created only an additional task for hub staff with no perceived benefit.
    The flag functionality was eventually removed altogether.
    
    In practice, the list itself is divided into two parts.
    New calls are added to the ‘Waiting’ section of the list.
    Once the ambulance service has assigned an ambulance (a conveying resource), the call is added to the ‘Active’ list.
    An individual call will be dynamically updated as new information is entered onto the system in the course of a call.
    Various alert icons provide information to those reviewing the list so that they can know if there has been a significant change in the events, information or status at scene or somewhere in the recorded detail.
        
\begin{figure*}
        \centering
        \includegraphics[width=\linewidth]{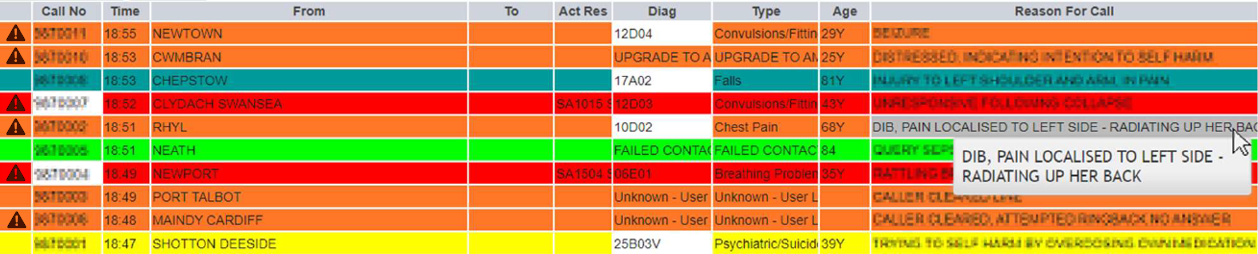}
        \caption[A section of the call list showing headlines of incoming calls]{A section of the call list (or `stack') showing headlines of incoming calls.}
        \label{fig:cad-stack-alerts}
    \end{figure*}

    In addition to a time, location and approximate patient age, incident detail is used to create a dispatch code (listed under ‘Diag’ in \cref{fig:cad-stack-alerts}.).
    Dispatch codes are determined by a series of question-and-answer protocols that cover different types of incident.
    At the beginning of a call, the call taker will choose one of 36 protocols and in so doing will select an ordered set of questions, the logic of which is designed to illicit the significant and salient facts from the caller about the type of incident being reported.
    In the course of the protocolised questions, information will be encoded about the nature and severity of the incident.
    The result is a five- or six-character alphanumeric code within the Advanced Medical Priority Dispatch System (AMPDS).
    The first two digits of the dispatch code (‘Diag’ column) indicates the protocol.
    And this can be displayed as a description column on the list (shown as ‘Type’ in \cref{fig:cad-stack-alerts}.).
    Each of the many thousands of AMPDS codes is deterministically assigned one of five main priority levels ranging from Red1 to Green3 according to local policy.
    Other colours are used to indicate events outside these priority levels (such as calls direct from healthcare professionals).
    Each list can be displayed in time order or priority order.
    And the user can set up the selection, order and size of columns to display.
    \Cref{fig:cad-stack-alerts} shows how a list is typically set up and illustrates the colours of the five main priorities.
    In addition to a code for the active conveying resource (‘Act Res’), the pictured list detail shows a simplified location ‘From’ column, a ‘To’ column and a short free text ‘Reason For Call’ entered by the call-taker.

    % \description{       
    % A tabluar list is shown with ten coloured rows visible. The rows are mostly amber with two reds, one yellow and one each of twd different greens. There is one symbol column and nine textual colums. The headings are, Call No, Time, From, To, Act Res (Active Resource), Diag (Diagnosis), Type, Age \& Reason For Call. From and To are location data for ambulance transfer (the To column is empty on new calls). Act Res shows a code for an ambulance vehicle and is mostly empty on new calls. Diag and Type indicate the chief complaint from a complex pick list. Age is in whole years (reported on the call). Reason For Call is very brief free text. A tooltip indicates that where this exceeds the column width, the user can hover over that field for the case and see the full text, which, in the example shown is still only two short lines describing breathing difficulty and pain localisation on the patient.
    % }   
    
    A major feature of work on the desk is that staff must ensure they are responsive to the overall dynamic status of multiple calls and at the same time be attentive to the detail within any that have their focus.
    Activity therefore continually switches between a review of new and changed calls on the list at headline level and concentration on an individual call in order to decide on whether to dig deeper in anticipation of a possible dispatch.
    This work is punctuated by dialogue which may be via the comms system or within the room - always aiming to home in on the pertinent detail that will help make an appropriate decision.
    A separate set of activities is always instigated whenever a critical care team is called into action.
    Once such a team is put into the field, they make use of the critical care hub as a communication and coordination centre.
    This means hub staff will facilitate case conferences with the top cover consultant - a senior medic not in the field who is available for clinical advice.
    But they will also provide an operational hub to minimise logistical conflicts such as those over resources or objectives between teams and vehicles.
    And they will liaise with receiving hospitals to ensure they are ready to accept the specific critical care case that is heading their way – by aircraft if necessary, and so involving transfer from a helipad and any special resources or timings needed for that to be successful.

\subsubsection{Informational profile of the process}
    \label{informational-profile-of-the-process}
    Before detailing our setting and findings from structured observation, we will review some important characteristics of decision-making in the different parts of the workflow.

    The informational context is that human decision-makers at the hub can access thousands of calls at a low level of detail (calls listed). 
    But they can only access a very small proportion at a high level of detail (viewed and interrogated). 
    This transition from a high-volume, low-detail context to a low-volume, high-detail context is what enables the process to achieve the high selectivity required over the initial input data. 
    
    The information at this first step (step A) consists of concise, semi-structured text.
    Dispatch coding, provided by call-takers, is standardised and hence straightforward. 
    It encodes a great deal of information on the type and severity of the incident using the AMPDS coding system mentioned above (\cref{ssec-features-of-the-call-list}).
    But the coding is not unambiguous.
    There are incidents that could logically fit under multiple different codes.
    And which one they end up being listed as will often depend on the order in which the caller identifies elements of the incident.
    For example, an incident that involves a fallen person underneath a heavy fallen object (or animal) is best triaged as a severe physical trauma.
    But it could well be coded as a fall - in which case, the protocol questions will, in turn, elicit details of the fall only. 
    So the AMPDS process could establish that the original fall was from a moderate or small height and encode that rather than the separate (but potentially more significant) risk of injury taking place immediately after the fall itself owing to the ensuing crush mechanism.
    
    Free text in the call listing can provide detail not captured in the AMPDS coding on which to make a decision to view.
    This is recorded using highly abbreviated and specialist vocabulary not always conforming to clinical terminologies. Most call takers are experienced, but not clinically trained. 
    And free text is rarely consistently rendered.
    It also varies widely in terms of the amount of information.
    As a result, the amount of relevant information in the headline listing varies between calls.
    Step A has high volume, low detail data.
    There is less information per call as compared to later in the workflow.
    
    Between steps A \& B of the workflow, the headline free text is augmented by a large amount of process data that arise from the call.
    This information is accessed by one or more mouse clicks and includes contemporary notes from the call-taker, geographical, timing and system information.
    Over half of all calls are reviewed by hub staff using this greater level of information.
    And as a result they are then able to make the decision (at step B) of whether to interrogate further.
    As mentioned, in a small number of cases, it is identified that interrogation is indicated.
    Importantly, staff may also identify an even smaller number where a dispatch is indicated immediately, without interrogation.
    This step (at B) has considerably lower incoming volume than is processed at A but each case has more detail to consider - a much higher level of detail per case.

    If the second step is passed (step B), a process of interrogation of the detail of the incident is initiated.
    Interrogation proceeds between steps B \& C and invariably involves dialogue with someone at the scene.
    This could be a first-responder who has already attended, a relative or witness, or even a dialogue direct with the patient.
    The channel of communication might be radio or telephone to scene.
    Or it might be that the hub staff can patch into the original incoming call as it is ongoing.
    This step is where many fewer cases have to be considered, but each has a lot of information which is highly complex.
    This is the highest level of detail per case.
    
    Such is the overall selectivity that those eventually dispatched represent less than 1\% of the original call list.
    At each step, the information is dynamic.
    Patients and situations can change rapidly and so a decision not to progress a case is contingent and subject to review.
    The informational enhancement of the selected calls comes at a cost of being able to interrogate a diminishing proportion of the total call list.
    The highest level of detail per case, achieved at step C, is on less than 2\% of all listed calls.

    \subsection{Results 2 - interactions between staff - and the detail of their situated action}  % NEW heading
    \label{ssec:interactions-between-staff}

    In \cref{ssec:immersion-methods} we noted the coordination of shared information between hub staff and the way the different interaction types were implicated in their work.
    By interaction types we mean not so much the technology used (radio, telephone, video, face-to-face) but the different communication functions.
    In this sense, we were interested not just in the \textit{how} of communication but specifically the \textit{who} and more especially the \textit{why} - what purpose in relation to the workflow did that interaction have?
    
    The answers to each of these questions is interdependent.
    Interactions with people in different roles will serve different purposes.
    And the communicative processes of dialogue entry, orientation, exchange and exit will differ, too, between different interlocutors.
    It is understandable and expected that human-human dialogue is structured quite differently between two colleagues who regularly work together and have been interacting in the same room for hours than between two others who are holding their first ever exchange over a telephone line.
    Each participant is aware of the existence and extent (or limits) of shared knowledge surrounding their interaction.
    This shared context is referenced by Suchman as the `mutually accessible world'~\citep[p 123]{suchman1985plans}.
    
    Awareness of what constitutes this mutually accessible world is clearly different within and outside the hub, but it is also different between the two colleagues, and recognising how differences are navigated is a key part of considering how decision support can work for a given task.
    In turn, understanding the work of establishing a shared awareness tells us a great deal about what work is being done at the critical care hub \textit{en route} to making dispatch decisions.

\subsubsection{Internal and external interaction}
\label{sssec:internal-and-external-interaction}
\begin{figure*}
        \centering
        \includegraphics[width=\linewidth]{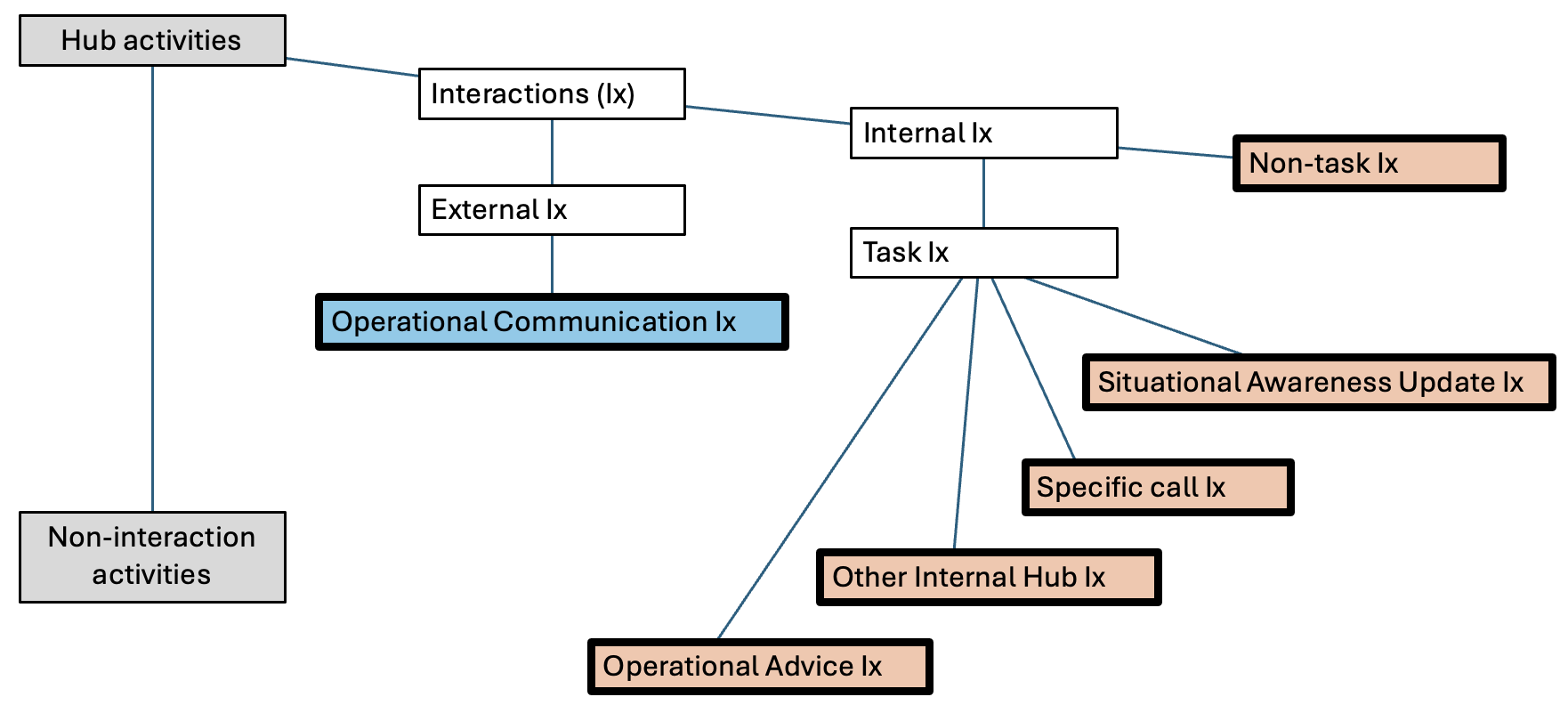}
        \caption[How six recorded groups of interactions arise from hub activities.]{How six recorded groups of interactions (Ix) arise from hub activities. The six groups we coded for are outlined in bold. Internal Ix are shown in brown, while external are in blue.}
        \label{fig:six-recorded-groups-of-interactions}
    \end{figure*}

    In \cref{ssec:observation-setting} we noted an interest in interactions between the two staff.
    And our detailed observations distinguish between these interactions, which we call internal, and those that are external to the hub.
    That is we count as internal those interactions between the two hub staff themselves.
    And we count as external all those where hub staff interact with others, whether face-to-face in the room or over communications systems out in the field or at other health facilities.
    \Cref{fig:six-recorded-groups-of-interactions} shows how interactions (as a subset of all hub activities) are split into internal and external interactions in this way.
    And our coding reflects the fact that the most common types of interaction overall are internal - those between the two colleagues themselves at the hub desk.
    We break these down into five groups according to purpose.
    Internal interactions include those that are not directly related to the work of the shift (non-task interactions), while most are task interactions and form the bulk of all interactions.
    Along with the operational communications of the external interaction group, we have six groups as described below, outlined in bold in \cref{fig:six-recorded-groups-of-interactions}, and summarised by frequency in \cref{tbl:obs-interaction-types}.

    	\input{tbl_obs_interaction_types}

    % \description{    
    % The figure is a hierarchical flow diagram showing how different types of interactions, abbreviated `Ix' arise from hub activities.
    
    % At the top left is a box labelled `Hub activities'. From this, two main branches extend:
    
    % a. Non-interaction activities — a grey box at the lower left.
    % b. Interactions (Ix) — a box near the upper centre.
    
    % The Interactions (Ix) branch divides into two categories:
    
    % First, External Ix, which leads downward to Operational Communication Ix.
    % Second, Internal Ix, which leads rightward to Task Ix and Non-task Ix.
    
    % The Non-task Ix category is one of the six recorded interaction groups.
        
    % The Task Ix category branches into four recorded groups:
    
    % Operational Advice Ix
    % Other Internal Hub Ix
    % Specific call Ix
    % Situational Awareness Update Ix
    
    % Thus, the figure identifies six recorded groups of interactions:
    
    % 1. Operational Communication Ix
    % 2. Operational Advice Ix
    % 3. Other Internal Hub Ix
    % 4. Specific call Ix
    % 5. Situational Awareness Update Ix
    % 6. Non-task Ix
    
    % The diagram uses colour to distinguish internal from external interactions: internal interactions are shown in light brown, while external interactions are shown in blue. The six recorded groups are additionally distinguished by thick black outlines.
    
    % In terms of structure, there is one external interaction group—Operational Communication—and five internal interaction groups: Operational Advice Ix, Other Internal Hub Ix, Specific call Ix, Situational Awareness Update Ix, and Non-task Ix.

    %     }

    While hub staff do spend a lot of time on external interactions, coordinating and providing or obtaining information from field teams, institutional health teams and other specialists, the interaction counts we recorded show that while this operational communication is the second most common of our six groups, it makes up less than a third of the observable interaction count.
    All other groups relate to interactions between the two staff.
    We discuss these internal interaction groups below.
    
    Updates that relate to situational awareness dominate all others in terms of occurrences.
    Situational awareness updates are those that don't relate to the internal details of a specific call (an individual incident) but instead establish the current status of active or available resources, the significance of several calls together, a resource conflict between two or more calls or a disambiguation between calls.
    The significance of these interactions is that such situational awareness forms the foundation on which many of the more case-specific decisions are made.

    Interactions between the two staff on specific calls constitute the third most populous group.
    In these interactions we identify the seeking and providing of initial and update details about an individual incident.
    We then have a group of interactions comprising utterances and actions that navigate joint tasks within the hub.
    These were coded as other internal hub interactions and were those not related to either situational awareness or specific calls.
    Notable instances are those in which assistance is offered or sought for some admin task.
    But the majority are responses of affirmation or negation.
    
    Although these short utterances may be responses to interactions in other groups, we nevertheless count these in the \textit{other hub} group because they are so numerous they would inflate the counts elsewhere.
    But they also tell us something useful - that a statement or query required only a one-word response.
    This is an indication that information was already held by the person prompting the reply.
    So this is a different kind of interaction to one in which data is lacking and the response must fill an informational gap with previously unencountered information.
    Again, the significance is in what this indicates about informational needs in the progress towards decision-making.
    Most of these are seeking a confirmation - a form of second opinion.

\subsubsection{Final grouping of coded types}
\label{sssec:final-grouping-of-coded-types}
    We actually coded five other interaction types within the last two groups, but have aggregeated them in our results into just the two groups without listing subordinate types.
    The first group is non-task interactions which included mostly organisational news such as how some service or staffing change would affect the work of the hub.
    But it included any housekeeping interactions between the two staff such as those that surrounded a break or the arrival of a visitor.
    These latter interactions, though not strictly operationally essential, are, in fact, essential to sustain communication since they lubricate and enable interactions that are more strictly `work-related'.
    Operational advice interactions form the last group.
    These were notable as rare moments when one member of staff was able to provide insight as a result of training or experience to which the other clearly didn't have prior access.
    These were instances not directly related to a decision to be made, but might be expanding on an option previously considered or discounted.
    They were provided as a form of professional knowledge sharing.
    Overall, our coding made use of twenty-seven interaction types that are spread over six groups.
    \Cref{tbl:obs-gps-ave-hrly-four-shifts} shows the average hourly count of each group over four shifts and shows both a consistency in the distribution of occurrences within the groups over the shifts alongside a difference between day and night shifts.

        \input{tbl_obs_groups_by_shift_ave_per_hr}

\subsubsection{Patterns detectable in the observation data}
\label{sssec:patterns-detectable-in-the-observable-data}
    During the night observations (which were the busier, early part of each night shift) the interaction occurrences were less frequent.
    Situational awareness updates were still seen to dominate each shift, forming a greater proportion.
    Operational communications also form a greater proportion of the total.
    But while these top two groups together accounted for an interaction approximately every two minutes on the day observations, they occurred closer to every three minutes on the night shift.
    Specific call interactions form a slightly smaller proportion of the count at night.
    And the minor groups are further attenuated, which was expected.
    The night shifts observed were ones in which remote desk working was undertaken.
    So the ease of (and motivation for) instigating some of these non-critical exchanges was seen to be lower.
    Another factor in reducing the overall night shift interactions is that the rate of incoming calls is reduced at night and only one of the four aircraft is equipped for night-time flying.
    So the level of rapid and intense triage is usually reduced.

    Operational communications feature interactions with ambulance crews, with receiving hospitals and with pre-hospital critical care teams.
    And these form the bulk of external interactions.
    It is worth noting that interactions with people on scene are often interrupted either by events or by some element of communications degradation.
    So these are frequently the most formalised interactions - involving acknowledgements, turn-taking signals, clipped grammar and explicitly repeated phrases of the kind that are associated with radio telecommunications.
    Important operational interactions occur with the Trauma Desk function - a trauma network specialist who sits adjacent to the hub desk and liaises with the network of specialist centres that take major trauma cases.
    This is a function that is external to the hub, but is covered overnight by the hub staff which increases their night-time communications load.
    The top cover consultant (TCC) is a senior, off-site, critical care medic who can provide detailed clinical governance, advice and decision-making around field use of advanced techniques and resource allocation.
    They are always arranged to be performing a non-operational role to ensure they can maintain objectivity away from the high cognitive and emotional pressure of events on-scene.

    \input{tbl_obs_types_by_shift_ave_per_hr}

\subsubsection{Unsolicited updates}
\label{sssec:unsolicited-updates}
    Our review of the captured data showed some interesting features.
    Some of the more detailed coding didn't translate across all observed shifts.
    We aggregated several of the twenty-seven codes as a result and this gave us fifteen types (see \cref{tbl:obs-typs-ave-hrly-four-shifts}).
    A notable feature of the work of the hub staff is that their direct communication with each other does not consist of a balance of seeking and giving information.
    There are many times more updates provided than are explicitly sought.
    This is most frequently in service of situational awareness.
    But it is also true for specific call updates as shown in \cref{tbl:obs-typs-ave-hrly-four-shifts}, while the same imbalance is not apparent in operational communications.
    So the anticipation element is a feature internal to the hub team.
    It reflects the close cognitive alignment of the hub staff indicating how each uses the other.
    In all three cases there are a variety of subjects of the interaction.
    For operational communications, the target of a query may be the status of the interlocutor.
    This is never the case for the colleague in the room.
    But otherwise, the range of subjects for each could include clinical or logistical information about a case, information about a resource, about the historical timeline, a predicted future timeline, or the effect on related decisions.

    Some of our original twenty-seven coded types did show up interesting patterns.
    We had codes that distinguished between solicited and unsolicited information-passing. 
    Taking the busiest hour for interactions on each of two shifts, we can see a marked difference in the approach to initiating \textit{within-hub} dialogue.
    Our data shows this to be reflected over the whole shift. 
    But it is most striking during these busy hours.
    And it is all the more remarkable because one half of the team in each of the two shifts was the same person - a very experienced hub allocator.
    The pattern (see \cref{tbl:obs-initiation-of-sa-two-shifts}) shows a marked contrast in the number of updates sought and provided in response, as against the proportion of unsolicited updates given.
    In one pattern the two staff explicitly initiate and sustain situation awareness updates (eg, the prospect of a given team to be clear for tasking). 
    In the other, they provide what is needed without a request being made.
    This contrast appears to be a function of adaptive working styles - an indication of variability inevitable among even a small number of humans.

\begin{table}[ht]
	% \centering
	% \small
	\caption[Initiation of Situation Awareness updates]{Initiation of Situation Awareness updates - comparison of the busiest hour from two different shifts}
	\begin{tabular}{l|r|r}
		\toprule
		Interaction Type & Count & Count\\
        & in hour & in hour\\ 
        & Shift A & shift B\\ 
		\midrule
        Sit Aw Upd - seek & 11 & 2\\
        Sit Aw Upd - give (solicited) & 14 & 2\\
        Sit Aw Upd - give (unsolicited) & 4 & 20\\
		\midrule
        Total & 29 & 24\\
		\bottomrule
	\end{tabular}
	\label{tbl:obs-initiation-of-sa-two-shifts}

\end{table}

\subsubsection{Factors affecting staff interactions}
    \label{sssec:factors-affecting-staff-interactions}
        
    The results illustrate qualitatively how variable each hour can be.
    But we also know, from prior work, that longer-run changes in the balance of tasks and the difficulty of decision-making can occur according to a number of factors in the wider service and the health community. 
    The protocolised question-and-answer system and associated training used by call-takers in the wider ambulance service is designed to yield consistent call records.
    Nevertheless, staff changes in the pool of call-takers can alter the level of detail and insight captured.
    In turn this can affect the ability of hub staff to identify details in candidate calls both on the list and during interrogation (when it is the call-taker who conducts the call and the hub staff listen in).
    Inevitably, there are variations in case mix as well as system pressures and this affects staff in roles across the service. 
    
    It is well known in England and Wales at the time of writing (2025) that ambulance services are struggling to respond to emergency calls within target response times~\citep{SummaryAmbulanceResponse} owing to delays handing over patients at hospital Emergency Departments as a result of staffing shortages and of chronic problems in social care.
    This has a significant knock-on effect on all emergency medical dispatch processes as the list of waiting calls builds up, call-taking staff try to support a higher number of what are now more agitated callers, experience higher levels of stress themselves and find that calls to scene can often be more fraught.
    Clinical field crews can come under extreme pressure as they experience the consequences of a higher number of cases that are exposed to what feels like preventable deterioration.
    
    In this context, there is also a real prospect that a specialist field team becomes committed to an incident scene and is unable to leave because there is no regular ambulance team to relieve them once any required critical care intervention is completed.
    So a specialist critical care team are left providing non-specialist care owing to system delays.
    Inevitably, this begins to affect hub thinking around dispatch decisions as the need to protect specialist services to be available for specialist needs comes to the fore.

\subsubsection{Variation over time}
\label{sssec:variation-over-time}
    Over the period of these observations, there was evidence of each of these effects, but obviously, not their long-run evolution.
    Changes that were evident were those resulting from the time of day and day of the week.
    A weekday that was extraordinarily busy was cause for comment from the hub staff. 
    A Saturday was not so comment-worthy as the staff expected a busy shift.
    Hour-by-hour, there were notable variations.
    Some of this was administratively driven.
    Prior to shift handover at 7am, there is a slow drip feed of new cases alongside a slow process of calls being resolved.
    The shift change is arranged to come ahead of an expected rise in call volumes. 
    
    But the hub staff also soon encounter the need to gather and disseminate daily informational briefings as hospital services come on-stream.
    As reduced night teams are replaced by full day teams in Emergency Departments, and clinical teams come on shift at air bases where there is no night flight capability, there is an organisational exercise akin to a \textit{roll-call} carried out through a sequence of dialogues conducted by the hub with each operational partner around the country.
    Hub staff compile a daily information set of who's fulfilling what role for the day, what resources are available and what the lasting effects of the night shift are for the day ahead.
    It is established who the Top Cover Consultant is and how they are contactable.
    There is a briefing of whether operating procedures are altered because of, for example, the unavailability of a major trauma unit and where patients should be routed as a result.
    This flurry of morning activity may give way to a busy morning for call interrogation or further activity with critical care teams in the field, depending on the level of activation of the teams.

    Another hour-by-hour qualitative change is that the balance between the number of critical care teams \textit{on-base} and those out in the field has a dramatic effect on what work the hub team are required to do.
    If all teams are \textit{on-base} then they are available for dispatch. 
    And the hub team are able to spend more time reviewing the call list and interrogating calls.
    If all teams are tasked to the field, then the coordination role of the hub staff can instead dominate their interactions completely.
    Staff must switch emphasis, but not lose sight of the list.
    So, at the same time, they will ensure that the call list is continually monitored. 
    At busy times, it is not impossible for a field crew to be moving from one call to a second and a third or more, without respite.
    The hub team, in these conditions can be supporting all four field teams attending multiple incidents over a period of several hours without break.
    All the while, they are monitoring the call list so they know how best to direct any crew that becomes clear to act.

\subsubsection{Sources of variation}
\label{sssec:sources-of-variation}
    Other changes that affect the interactions of hub staff include weather conditions across the country and whether the hub staff are co-located or not.
    Weather conditions can, of course, be highly localised. 
    And the decision of whether flying is an option or not is taken by the aircraft pilots.
    However, the hub staff seek to anticipate change in availability as conditions change.
    So there can be a lot of dialogue with local teams to check on how things are evolving.
    A deterioration in weather conditions may mean a Rapid Response Vehicle is used instead.
    This, in turn, means there are different logistics around travel times and getting to scene.
    An aircraft is faster over long distance, but invariably entails some kind of transfer plan since it cannot fit into the small footprint of a road vehicle and so will often need to land many metres from the scene of an incident.
    The presence of trees or overhead power lines may often mean a helicopter landing cannot be as close as could be achieved by road.
    It also alters the logistics around patient transport since not all patients are suitable for air travel while the critical care road vehicles cannot convey patients.
    Each change to external conditions has a counterpart in changes to the work of the hub staff.

    The co-location of hub staff has historically always been standard practice.
    But following technological developments in response to the covid pandemic, it has become possible for the team to operate using the same systems from different locations, and this places less of a travel burden on staff doing what is quite a punishing shift pattern.
    So a proportion of shifts are now carried out where the two hub staff are working from different locations.
    The two distinct features of remote hub working are that communications between the two hub staff have to be initiated explicitly and that their own dialogues with third parties are frequently muted.
    
    Taking each in turn, we noticed that co-located working allows hub staff to initiate communication non-verbally and without an explicit request.
    For instance, the researcher observed several instances of one of the staff attracting attention of the other by non-verbal means in order to not interrupt an ongoing dialogue on an external line.
    Hand signals and pointing at information followed in answer to the evident query.
    We also saw that each of the hub staff could stand up and move to the other's workstation enabling them to look at exactly the same information with minimal signals passing between them.
    
    For the muting behaviour, the researcher had noted for co-located staff that there were frequent times when only one was on a call, but the other would be able to contribute by dint of knowing which case was under discussion and hearing one side of the exchange.
    When observing remote desk working, the muting of the shared channel when on an external call line made it easier for the researcher to concentrate clearly on the one speaker. 
    But there were many occasions where dialogue had to be reported following unmuting after a third-party call, so situational awareness updates were more explicit as each person was unable to rely on incidental or deliberate listening in by the other.
    This does not account for the different patterning of solicited and unsolicited updates noted in \cref{sssec:unsolicited-updates} above as remote working was not a factor in those patterns.
    An additional factor in remote working is that there were several instances of online communication breaking down when the audio dropped.
    And this precipitated conversational repair events that were extremely rare for co-located staff.
    
\subsubsection{Shared information}
\label{sssec:shared-information}
    There appeared to be challenges on occasion for the researcher in identifying what counted as giving and seeking information.
    The initial assumption was that unsolicited information-giving would tell us something about the working relationship between the two hub staff.
    But further consideration suggested that it was their joint immersion in the incoming flow of information that led to the frequent dispensing with the commonplace conversational pointers to a new topic.
    Without a demonstrative, deixis or identification, it appeared as though they were regularly engaging in non-sequitur.
    But this impression abated as it became easy to recognise the ‘informational context’ shared by the two staff.
    Although the two staff might sometimes be aware of identical information, it is obvious that their attention to the call list and the mass of detail behind each item usually differed.
    Indeed, again, as evident from their dialogue, they were in a constant state of updating and checking their common pool of knowledge about cases gaining their attention.
    In this context, an utterance that is not preceded by an obvious social or conversational prompt is simply an expression of the common task.
    In other words, the occurrence of unsolicited utterances is less a reflection of staff behaviours and more a reflection of the expectation of a common focus of attention within the ever-changing ‘informational landscape’ being revealed to the team minute-by-minute.

\subsubsection{Situated action - simplifying a complex process}
\label{sssec:situated-action-simplifying-a-complex-process}
    Having used structured observation to obtain considerable detail around the interactions between the two staff at the hub, between them and colleagues and between all of these people and the information they encounter, we are now in a position to review how our three key decision-steps fit into the overall dispatch decision process.

\begin{figure*}
        \centering
        \includegraphics[width=\linewidth]{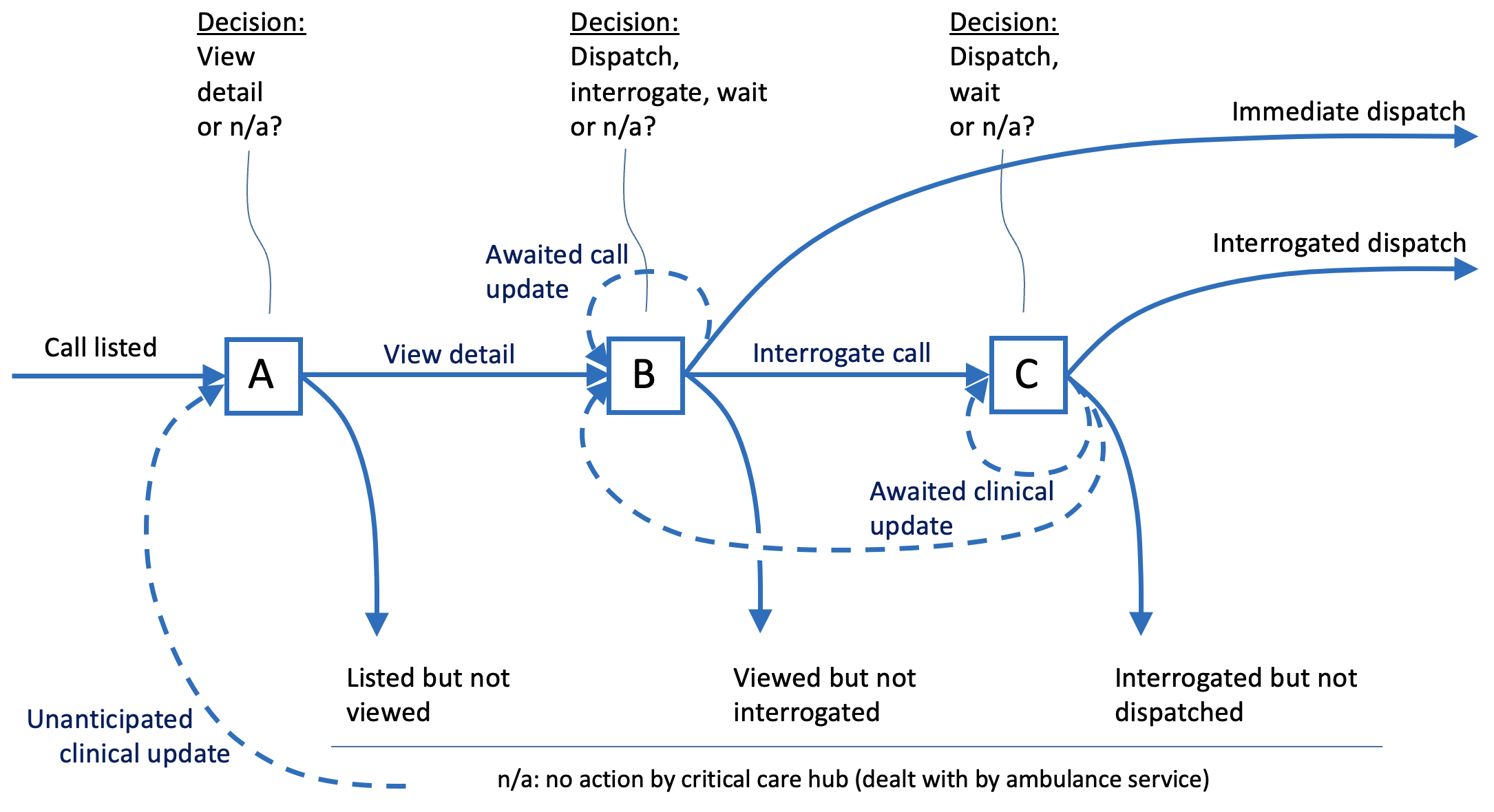}
        \caption{Three key decision points identified in the work of the critical care hub}
        \label{fig:ecch-three-key-decision-points}
    \end{figure*}

    Having shown the overall process in outline in \cref{fig:ecch-three-decision-steps-outline}, the more complete process with its three key decision points and two kinds of dispatch outcome is shown in \cref{fig:ecch-three-key-decision-points}.

    We can see that the whole workflow is complex.
    The usual process leading to dispatch consists of two triage decisions (A and B) followed by a substantive dispatch decision (C).
    In reality, there are many more decision points that we have left out of consideration to reach this simplified view. 
    Even prior to a decision to view detail (A), there is a decision to attend to the list itself. 
    At a busy desk, time must be made to do this.
    At any given time, there may be competing demands, such as the need to try again to get through to hospital X or team Y who were un-contactable before for some reason.
    Or there might be an important (while not urgent) need to document earlier events in a patient care record. 
    So there is always a precursor to decision A in the form of a decision to attend to the list itself.

    Following a decision to open a call to view its detail, the decision (at B) will rest largely on what is found within that call detail.
    But events outside that call may overtake the process so that viewing is aborted - a different decision is made.
    Within the viewing process, there is also a continual decision process taking place on whether to read further or stop viewing.
    A form of exploration-exploitation trade-off~\citep{bergertal2014ExplorationExploitation}.
    Before making any substantive dispatch decision (at C), the hub staff must also take into account whether (and if so which) resources are available.
    There may be previously-advised information from a given air base that operations there are impacted by weather or technical developments for the next several hours.
    Or a crew on scene may update with an unexpected change in their mission's clinical objectives.
    It may be necessary to get an update on such resource constraints before making a decision on the current case.

    The retrograde loops (shown dashed in \cref{fig:ecch-three-key-decision-points}) increase the complexity of the decision-process.
    In each case, we can see that the informational input may arise, not from the immediate predecessor in the normal sequence, but from a separate process of seeing or hearing update information - whether concerning the call or the resources.
    So there is a class of decisions concerned with attending to, seeking out and possibly prioritising update signals.
    And while hub staff are experienced in identifying calls that are likely to change with an expected update (ie, those worth `keeping an eye on'), there will always be cases where an initial decision was made for no action by the hub team (indicated as `n/a' in \cref{fig:ecch-three-key-decision-points}) but where subsequent events will create an update on the list and this could constitute an unanticipated clinical update requiring re-consideration of the call at decision-step A.
    Any given call may be initially coded a lower priority (eg, Green2) but then upgraded to a higher priority (such as Amber1).

\begin{figure*}
        \centering
        \includegraphics[width=\linewidth]{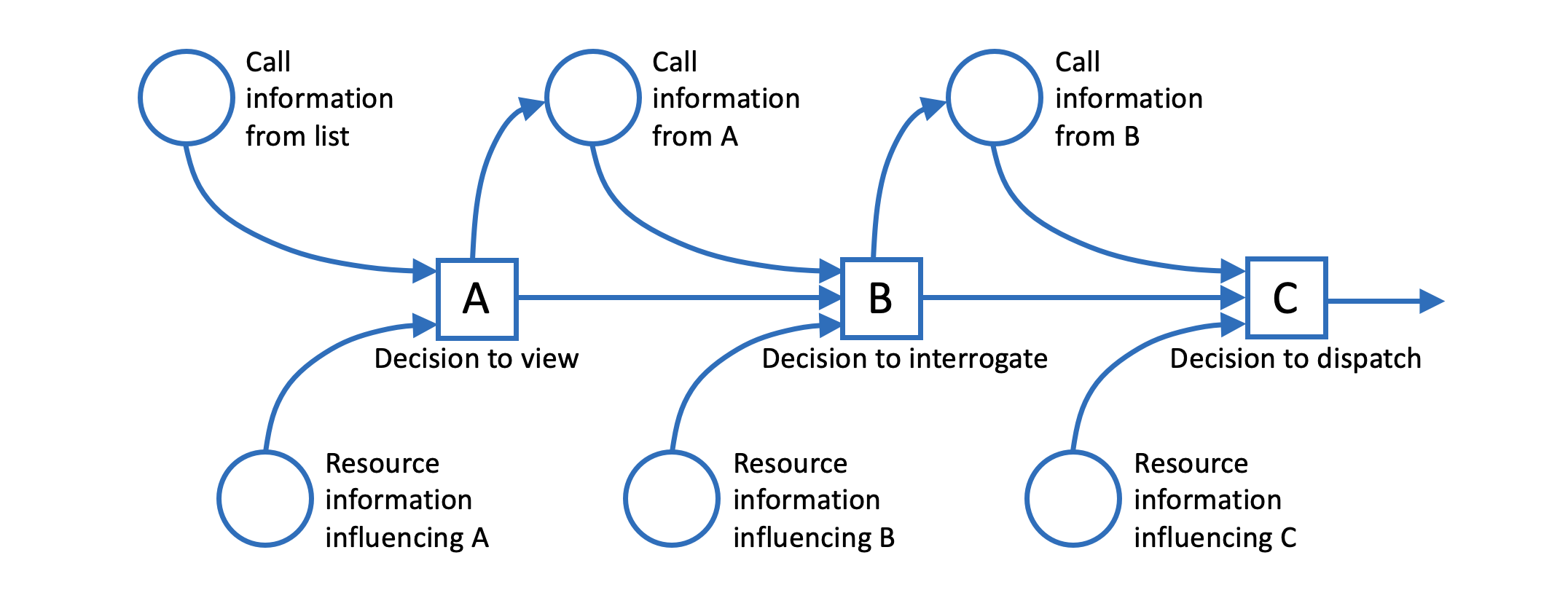}
        \caption{Influence diagram for the three key decision points}
        \label{fig:ecch-influence-diagram}
    \end{figure*}

%\subsection{star section moved from pp86-87 to after Amber1 on p88}
    As mentioned, representing the process as three key decisions is a simplification.
    But as \cref{fig:ecch-influence-diagram} shows, the information acquired ahead of each of these steps constitutes new influence on each ensuing decision - so decision theory requires us to recognise these three steps as distinct decisions at least~\citep{howard1968FoundationsDecisionAnalysis}.
    Case information at A comes from a single line of information on the call list.
    At B, this has been augmented by call information obtained by opening the call to see detail of what was recorded during the protocolised interaction with the call taker.
    And at C call information has been supplemented by audio - what is gleaned from interrogating the call - or even by video, if the on-scene video app (GoodSam) is called into use with the help of a bystander or first responder.
    The resources and team dispostion are also inevitably different at each step.
    At quieter times, a greater number of the less acute cases will be included in those opened in step A.
    At step B, the relative location of the incident and any available team might determine whether one or other borderline case is interrogated.
    Of huge significance at C is whether there is a team available within a critical timeframe.
%\subsection{end of star section}

    Despite these complexities, we justify our simplification on the basis that each of the three steps shown represents a definite point in the workflow where information available to the decision-maker has necessarily changed.
    
    The whole process can be seen as a complex layering of multi-step, multi-source, triage processes that converge on decisions for action.
    And this is before we consider that decisions for action may well be to intervene for the purposes of providing remote critical care advice without any prospect of an actual critical care dispatch itself.

    Given the reality of all these multiple decisions, then, how can we justify this simplification to three key steps?
    We do so by making several distinctions.
    We can distinguish between the clinical assessment of whether a call could benefit from critical care intervention (regardless of resources) and the operational assessment of whether the service can actually provide support in this case and this location (given the resources actually available).
    In practice, hub staff are always dynamically considering how the disposition of their field teams affects the acuity of their next candidate for dispatch, so these processes are not separated. 
    But they must anticipate the possibility that a given field team may become available sooner than expected.
    So at busy times, the staff will have an idea of what might be next in line by considering the clinical need even if practicalities look like preventing action.
    
    We can also distinguish between the core decision pathway and how it is distorted by interruptions and events on the ground.
    The core decisions shown in \cref{fig:ecch-three-key-decision-points} are made on every interrogated dispatch.
    Ancillary decisions may or may not be implicated.
    With these distinctions, we can see the value in isolating the key decision steps and focusing on how these begin with the call list and end in a dispatch or a decision for no further action by the hub itself.
    It is along this route that the key decisions affecting a dispatch are made.
    And it is therefore these key steps, or their combination, that become the candidates for the kind of algorithmic support we wish to explore.
    Putting this description of the key decision workflow alongside the structured observation findings about shared information allows us to see how dynamically the workflow builds.
    We next review the prospects for algorithmic input into this situation.

\subsection{Results 3 - locating where to assist}
    % \subsubsection{Reviewing the key steps as prospects for algorithmic input}
    % \label{gemba:sec-reviewing-prospects-for-alg-input}
\label{ssec:locating-where-to-assist}

    Anticipating the need to consider an algorithmic assist, we mention here some accumulated requirements any AI system must meet -- and some connections between this work and evaluationki work at the next stage.
    To be effective, machine-learning-driven decision support in these contexts must be clinically robust. 
    But it must also allow humans to continue to function in a highly interdependent workplace where multiple processes are carrying on in parallel, task-switching is frequent and situational awareness must be rapidly and repeatedly acquired, maintained and re-acquired. 
    Aspects of situated human-computer interaction (HCI) are therefore as critical as the level of algorithmic performance.
    Moreover, what should count as worthwhile algorithmic performance and how it should be evaluated is dependent on how humans can access any algorithmic contribution.
    Involving clinically trained and experienced critical care practitioners in the project will provide an important level of ecological validity to any subsequent experimental work that is informed by these findings.
    But the fact that their later-proposed empirical contributions are to be based on experimental design that is informed by a close study of their situated work also means that their engagement and confidence in the value of the work is likely to be maximised.
    
    The work of these specialist practitioners - human decision-makers - at the critical care hub is our main focus.
    Our aim, then, is to consider the context within which an algorithmic assist tool needs to operate - the context within which it needs to be of benefit and within which evaluation will need to take place.

    Because of its high selectivity, and unlike most emergency medical triage, the process leading to a potential dispatch in pre-hospital critical care has a distinct structure.
    Understanding this process is essential in order to identify where and how algorithmic support can be considered.
    
    As can be seen, the specialist workflow sits over the existing emergency ambulance process and provides a small number of highly specialised interventions.
    So, the pre-hospital critical care dispatch process has features in common with more general medical decision-making. 
    A first decision might be on whether to seek more information. 
    But the value of that decision may be determined by what information it can bring to a subsequent definitive decision on care. 
    In a sequence of such decisions, there is the potential to encounter diminishing returns - so that more information costs time, resources and patient discomfort (or even risk) without adding significantly to the definitive decision. 
    For the individual staff member, all this must go on whilst there are multiple other tasks demanding attention.

    A significant difference between human and algorithmic decision-making is the speed and volume of information processing that computer algorithms can achieve - leveraging the information processing power of machines can often mean that even weak classifiers can provide benefit in terms of alleviating cognitive burden.
    Where in the process we can make best use of algorithmic capabilities is a key design question.
    In addition, since decision-making must go on whilst there are multiple other tasks demanding attention, algorithmic decision support needs to be provided at the right point in the multi-step process with timeliness and appropriate subtlety to avoid distracting the human user from the multiple tasks in hand. 

%\todo{moved informational trade-offs to before obs}

\subsubsection{Three candidate steps for decision support}
\label{sssec:three-candidate-steps-for-decision-support}
    A first question is whether an algorithmic process could accomplish all of this multi-step decision-making and provide a ready-made list of suggested dispatches for the human.
    Our discussions with staff suggested a reluctance on their part to have an algorithm provide suggestions directly on dispatch itself.
    The perspective is that human oversight would only be meaningful if the human in the process had been able to review the interrogated call information.
    Or, at a minimum, the algorithmic assist system would need to provide both justification (reasoning) and evidence (source data) from a commentator at-scene who was reliably identified as falling into one of at least six principal roles in order to contextualise their information (eg, patient, first-responder, paramedic, bystander etc) or compelling content from the sequence of events record or call history justifying an immediate dispatch.
    In this sense, potential users wanted  to see a large informational overlap between themselves and the algorithm. 
    And they would want to see information on the intermediate steps they themselves would have navigated to reach the suggested decision.
    
    Relevant here is that staff reported that their sense of what they bring to each step is informed by their own clinical experience.
    Reference was often made to notable cases from their direct experience in the field that they identified as learning points.
    These were usually where the outcome or course of clinical care differed widely from what they had initially expected.
    Importantly, the pervasive culture maintained in the team is to reflect on such cases regularly and try and identify reasons for any initial misdirection and to always evaluate and share useful learning from the process.
    This was evidenced not only in dedicated clinical governance meetings, but in frequently-witnessable incidental discussions over coffee or collegiate news-sharing, frequent `checking-in' with each other and the provision of mutual support.
    So an algorithm would have to demonstrate incredibly high reliability and be able to adapt to (or else persuade) individual clinicians of a line of clinical reasoning to stand any chance of acceptance as a source of ready-made dispatch suggestions.
    
\begin{figure*}
    	\centering
    	\includegraphics[width=1.0\textwidth]{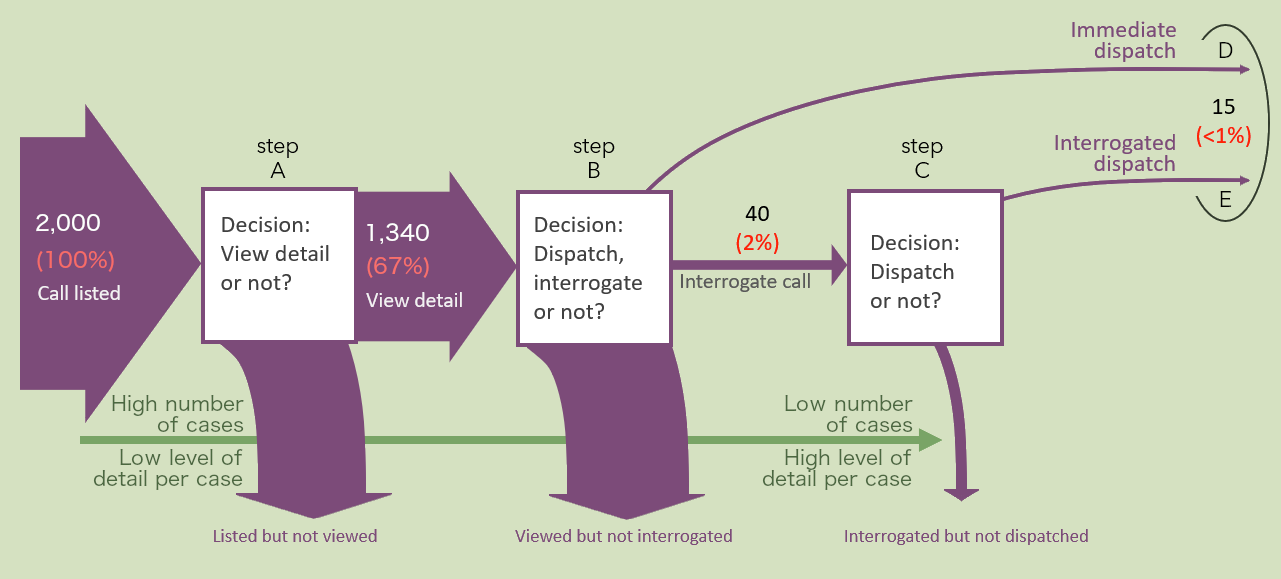}
            \caption[The flow of calls through the hub]{The flow of calls through the critical care hub workflow.\\
    	\label{fig:ecch-workflow}
    	}
    \end{figure*}

    At the same time, looking at our three decision steps with an indication of flow volumes and information density (\cref{fig:ecch-workflow}), high levels of informational detail per case (as are encountered at step C) tend to imply greater algorithmic opportunity for machine-based techniques.
    But the availability of information in an organisation doesn't mean its ingestion into a machine pipeline is straightforward.
    While the information at the first triage step (step A) is indeed scant, it is relatively homogeneous and it is readily available in a single electronic format and within a single information system. 
    Subsequent decision steps recruit information from multiple sources and systems in highly heterogeneous ways.
    And the information is multimodal.
    Some of the information a final decision is based on will have come from scene – usually in the form of an audio stream - sometimes in the form of video.

    Dispatch decisions are always influenced by whether a team is available to act. 
    This can quite rightly mean they are impacted by environmental conditions such as time of day, wind speeds, cloud levels, visibility etc. - all information that is available in disparate systems and is geographically expansive.
    But such information is only relevant in relation to the route needed.
    So it is not all pertinent.
    More than this, if all teams are busy at scene or \textit{en route}, then decision-making at the hub has to shift to accommodate the temporary lack of resources. 
    This information on the disposition of field teams is communicated regularly, but currently stored only in human memory - along with experienced predictions of when they are most likely to be clear for the next task.
    
    All of this taken together means that if we want to model the third step, we would need to ingest data from a diverse array of systems - and introduce and populate some new ones that don't currently exist.
    Commissioning, routing, combining and maintaining these information sources in a unified and sufficiently robust data pipeline would be a considerable undertaking.
    For it to produce live data feeds so as to be useful in real time would be very costly.\footnote{Organisational information systems are interdependent and the modification or replacement of a single one is a long-term project requiring extensive risk assessment and mitigation. The assemblage of systems existing at any one time in a large organisation, such as an ambulance service is usually the result of historic development that prioritises resilience - a development involving tailoring system outputs so they can be ingested into existing critical operational data functions. Adding extra data feeds or altering existing ones usually requires a high level justification. Substantive system changes usually require a full business case.}
    This makes the information at step C, in particular, a candidate only for a very long-term machine-based data modelling programme.
    Decisions to view and interrogate, in contrast, are taken on information already sitting within the call listing system itself. 

    As against this, modelling decisions made on the relatively low-density (albeit high-volume) information that is streaming through the call stack represents a significant challenge. 
    But is it a more tractable problem than modelling decisions on dispatch itself?

    The fact that system developers have shown enormous ingenuity in bringing together diverse sources of multi-modal data raises the question of whether the ambitious target of addressing step C might be worthwhile.
    Discussions with staff, however, encountered potentially more significant problems with considering decision support at step B or step C.

    \subsubsection{Interaction with an algorithm when time is tight?}
    \label{sssec:interaction-with-an-algorithm-when-time-is-tight}
    
        The problem of review would be particularly acute with any algorithmic tool that wasn't capable of rapid clinical and situational interaction or exchange.
        One user justified this concern by pointing out that they could ask a colleague why they made a particular judgement.
        And what they'd get back would be extremely concise and it would be simple to probe and challenge with little chance of needing to re-word their questions.
        Algorithmic justifications, by contrast, were imagined to be verbose and disconnected - with questions to the algorithm likely to be difficult to formulate, too easily misunderstood and too readily requiring a crafting process.
        
        At root, an algorithm's involvement would make the clinician want to double check what information it had picked up, and from where.
        But that process might involve yielding control of the screen to the algorithm while it either displayed an extracted part of the sequence of events log or highlighted a piece of text in its wider context or (horror!) even scrolled the user's own view unexpectedly.
        Crucially, the amount of context required for an individual point to become meaningful would depend on the case and the user.
        Any assistive process could easily risk becoming too intrusive unless extremely high-performing for both.
        As a result, users felt algorithmic support would not be suited to the later decision steps.
            
    \subsubsection{Limits to algorithmic situational awareness}
    \label{sssec:limits-to-algorithmic-situational-awareness}
    
        Further probing on many of these issues brought the question to the fore of whether the focus on decisions-to-view was justified or whether it might be more effective to apply algorithmic techniques to later decision steps in the workflow.
        
        One reason given for scepticism about the involvement of algorithmic tools in the later step of dispatch decisions was that the clinicians at the hub have a detailed knowledge of the skills of the various types of ambulance crews at, or close to, scene.
        So they habitually tailor their decisions - as well as their advice and questions to the specific crew - according to what they know of the skill-set and equipment available.
        An algorithmic approach would struggle to know these things reliably owing to the fact that documented crew resources and real resources are frequently at variance.
        And sources of good, real-time information are usually in people's heads rather than in systems.
    
        In a similar vein, a human dispatcher may be thinking several steps ahead, for example, in relation to which major trauma centre would be best for treatment.
        Which receiving hospital had limited capacity today?
        Which critical care team was already struggling with an equipment problem at the start of shift?
        Capacity at different centres and on the different teams changes daily (and can sometimes change hourly), with information gathered by hub staff at the start of their shift and updated throughout the 12 hours.
        Such information is stored, essentially, in people's heads or on paper, not in a daily database.
        Having to feed such information into a machine every time the situation changed would make it prohibitively burdensome.
        And it would reduce human attention to clinical decisions.

    \subsubsection{Algorithmic assistance in the workflow}
    \label{sssec:algorithmic-assistance-in-the-workflow}
        
        A common theme in clinicians' views was that they didn't feel confident that an algorithm could bring clinical judgement to a decision to dispatch.
        During interrogation of calls (when a dispatch decision would usually be made), it was pointed out that there is a `whole picture' to be evaluated. 
        `It's not just the information but the context of that information', said one participant.
        More than one user described how sounds in the background of a call might be signals that would need interpretation and nearly always specific questions to the person on scene.
        The sound of a baby crying on a paediatric choking call would immediately prompt an uncertain clinician to ask if that was the patient.
        In a similar way, a human can ask for, and interpret, information that reveals the kind of mechanism of injury involved in a road traffic accident by asking how far away the cyclist ended up, for example, or whether there is a dent in the van or about the nature of a cracked windscreen.
        Hub staff did not believe an algorithm could play that interactive role in real time.
        Asking clarifying questions is an important part of interrogation, even if the clinician is inserting those questions into the dialogue for an ambulance service call taker to voice.
        On the whole, there was very strong agreement that algorithmic assistance should be confined to the early parts of the workflow.

\subsubsection{The learning problem - evaluating decisions in a dynamic context}
    \label{sssec:evaluating-decisions-in-a-dynamic-context}
    
    At each triage step, there is a complication owing to the varying conditions under which trade-offs have to be made.
    Exploration ahead of a triage decision always involves diminishing returns.
    More time spent gathering data can help improve the decision.
    But only up to a point.
    The time used in viewing or interrogating one case reduces the time available for exploring others.
    This means time available for steps A and B is a function of all the other competing calls.
    At busy times, fewer calls can be viewed or interrogated.
    Less busy times mean more calls can be explored or the same quantity explored in more detail ahead of each triage decision.
    Even with theoretically perfect evaluation, labelling could only ever apply a relative weight to each case as the available time to make that evaluation is dependent on how other cases compete for attention.
    It's impossible to establish a best binary decision outside the dynamic context of time pressure.
    Even a ranking of cases would not be static since externalities impinge on cases differently.
    
    At steps A and C the decisions are relatively well-formulated. 
    In each case, if we ignore the question of whether to wait for updates, there are two possible routes for a case to take. 
    So the outcomes are binary.
    The decision at step B is distinctive in that there are three rather than two immediate outcomes. 
    As a result the required learning process for an algorithm to operate at this point is vastly more complicated.
    Even if the compound decision were separated into two parts, it still represents a considerable challenge.
    The first part would need to determine whether an immediate dispatch was justified.
    And if not, a second part would decide whether interrogation was warranted.
    But there is no clear definition of `immediate dispatch' in practice.
    The category arises in the course of narrative guidance that reflects only marginally on the `work as done'.
    Hub practitioners allow for a highly variable amount of interrogation in recognition of the individual nature of each call.
    
    As in any clinical context, triage decisions here are finely balanced.
    In the process as a whole (as well as at each step) – any move to address over-triage (trying to increase specificity) is likely to increase under-triaged cases (reduce sensitivity).

    In most decision-making, there is a recognisable benefit in gathering more information.
    The effort expended, however, must always be considered against the influence new information can have in improving the decision.
    As mentioned, it is a diminishing return. 
    In other words, there is always a less visible cost.
    The hub staff never have perfect information.
    But they always have competing cases.
    So a decision to seek further information on one case is in itself a decision - with a clear benefit but with an often-hidden cost, that is, not reviewing other cases. 

    Several hub staff report multi-tasking - that is, they may listen in on one call, participate in discussion on another call and review yet other calls on the call list all at the same time.
    Even with this high level of parallel working, the call list can be reviewed only one case at a time.
    So, for the first step in the hub workflow, the time available for viewing detail is a function of how many calls there are to review.
    At the beginning of each shift, most hub staff aim to review as much of the call list as possible.
    On `quiet days', this is achievable.
    And the task is aided by the handover process between shifts.
    During the course of the shift, after each period of having `eyes off' the list, each user aims to re-acquire a situational awareness of the updated list.
    But if the rate of incoming new calls is high, or a new coordination task constrains their review, then the user must either make an incomplete review or else must make their decisions to view at a faster rate.
    In other words, the threshold determining whether a call is viewed in detail is not a fixed one.
    At quiet times, hub staff can afford to view the detail of more calls.
    At busy times, they can afford to view detail on fewer calls.

    All this has implications for the process of establishing good learning data.
    What makes a good decision to view or a good decision to interrogate depends not only on the case in question, but the competing calls for the attention of the hub staff.
    In other words, the value of the information gained from a detailed view is not only impossible to know in advance, but is impossible even to fix post-hoc.
    The best that can be attempted is to apply a relative weight to each call indicating the relative value of it being reviewed in what must be an estimated average background list context.

    There is a challenge arising from the incompleteness of early data and the multi-modularity and multi-component criticality of later data.
    It is a challenge compounded by the difficulty of investigating counterfactuals.
    That is, the great difficulties involved in `ground-truthing practices'~\citep{jaton2021AssessingBiasesRelaxinga, cabitza2023PerspectivistTurnGround} are compounded by a situation where real-world results come only after three decisions have been taken at the hub and several subsequent decisions and interventions have been carried out (or not) at scene.
    Resolving the positive (or negative) contribution of each decision is fraught with contextual and counterfactual caveats.
    
\begin{figure*}
    	\centering
    	\includegraphics[width=1.0\textwidth]{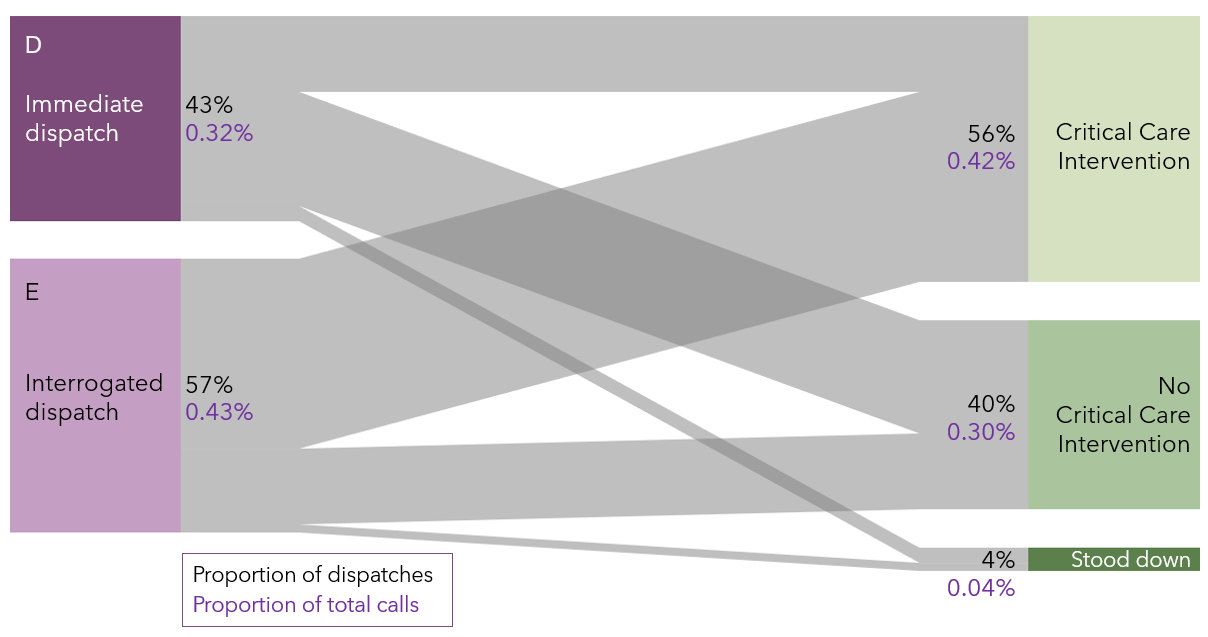}
            \caption[Sankey diagram of flows following dispatch by the hub]{Sankey diagram showing flows following dispatch by the critical care hub.
    	\label{fig:sankey-diagram-of-call-flows}
    	}
    \end{figure*}

    The disposition of call flows at the hub are shown in \cref{fig:sankey-diagram-of-call-flows}.
    It is clear that the data of fifteen dispatched calls out of 2,000 is unlikely to provide enough relevant information to help improve the large mass of initial triage decisions - even assuming we had selected the optimal fifteen.
    And, as we have pointed out in \cref{ssec:cch-workflow-in-outline}, how busy the hub and the field teams are on a given day will determine whether a decision to view is appropriate or not.
    So we are unable to pin down an objective `ground truth' for any of the decision steps based on historic clinical outcomes data.
    Learning data for the decisions at each step must build on what we can know about the immediate outcomes of those points in the workflow.
    The human expertise currently in the hub team is the best source of evaluation data.
    But how this expertise can be extracted and harnessed is at the heart of the improvement problem.
    So our concern with the feasibility of frequent and regular, low-cost human feedback has heightened significance. 

    We had access to anonymised historic data allowing us to see the immediate organisational outcomes.
    That is, whether an actual critical care intervention was made in the course of a dispatched and attended call.
    The flow diagram (\cref{fig:sankey-diagram-of-call-flows}) shows the different outcomes of the two kinds of dispatch decisions we saw above.
    The exit flows at the upper right of \cref{fig:ecch-workflow} are the two inputs on the left here.
    Where there is a critical care intervention, it means critical care clinicians provided their specialist care out in the field as a result of the tasking.
    What it tells us is that the patient was in a condition to benefit from the specific care that only these teams are able to deliver.
    Where there is an attendance at scene but no critical care intervention - or where the crew were stood down before arriving on scene - it tells us that the patient was not in a condition to benefit.
    
    It's tempting to argue that this information is the best test of whether the decision to dispatch was a good one.
    And we can know this information, because we can find out from clinical records whether a critical care intervention was made. 
    But, at best, this gives us a one-sided view of the process. 
    If a patient who was judged a good candidate for critical care sadly deteriorates and dies before an intervention is possible, they will end up counted here as having had no intervention. 
    It’s unlikely that we can conclude that in every such case the dispatch decision was poor - a false positive decision. 
    The informational uncertainty is too great to make that reasonable. 
    By definition, the decision to dispatch is made on different information to the decision to instigate critical care.

    There is also a patient care benefit from the presence of critical care cover during conveyance to a hospital Emergency Department.
    Some patients may be judged in need of such close and specialist monitoring.
    And then providing qualified staff to cover the journey is a justifiable use of specialists.
    But it may be that no critical care intervention is made, and no clinical coding will capture the fact that a case represented a justified use of the resource, and so was an appropriate dispatch.
    In the recorded data, it will be indistinguishable from a record where it turned out that an inappropriate dispatch was made.
    
    At the same time, in routinely recorded service data, it is difficult to see the false negatives - where no dispatch was made, but critical care might have improved the outcome.
    We have to remember that the call list is an order of magnitude larger than the two dispatch volumes on the left of \cref{fig:sankey-diagram-of-call-flows} combined.
    Together, they make up less than one percent of calls in the overall list. 
    We get some idea from subsequent hospital notes, but a reliable set of false negatives to learn from would be difficult and costly to assemble.

    % It isn’t recorded at what point in time the decision about the level of intervention was made. 
    
    Two cases with otherwise identical indications at the time of dispatch can lead to very different decisions on scene for a plethora of different reasons.
    The outcome of either one may not teach us anything about the quality of the decision made on either case.
    And this is primarily owing to the impossibility of improving the information at the time the decision must be made.
    
    There is a natural sense in which the decisions at step B are more difficult than those at step C since the easier-to-determine cases have been eliminated.
    At the same time, there is an imbalance in error costs between the steps.
    At all steps, the cost of a false negative is equally high - denial of critical care to a patient who could benefit from it.
    And overall, to minimise the risk of missing a suitable case, hub staff must trade off specificity for sensitivity.
    But the time cost of a decision at step B is much more significant than at step A.
    Each requires more work at B. 
    So the price for false positives arising from step A are paid at step B -- which is a consideration suggesting that step B might be a less suitable place for initial experimentation.

    Looking beyond initial model training, and assuming we would want to be able to improve - or even merely evaluate - any deployed system, there would be significant benefit from being able to provide efficient feedback on its output.
    And this would be an important step towards enabling sufficient and demonstrable, situated reliability as well as guarding against data drift or problems created by inevitable software updates to source systems.
    Feedback on a small sample of cases each shift where the clinician view diverged from the machine suggestion is conceivable at step A.
    But comparable feedback at step B would be time-consuming and probably impractical, while at step C it would undoubtedly prove prohibitive.
    A separate audit and improvement process would need to be created, funded managed and evaluated.
    The argument for decision support being targeted at the earlier, triage steps is gathering strength.

\subsubsection{Support for situated decision-making}
    \label{sssec:support-for-situated-decision-making}

    Looking at how decision-making is carried out is always valuable when considering how it can be supported.
    For hub staff the process involves an attempt to build the clearest picture of clinical signs and likely prognoses.
    But in building this picture, the process of gathering information from scene is itself rarely straightforward.
    And, as mentioned, the situation at incident sites is always subject to change - either as a result of a clinical change on the ground or because information is newly available to the critical care hub.
    So there is frequently an array of calls on the list that are of simultaneous concern to the hub team.
    Ambulance service teams may also consult with the hub clinicians in the course of an incident involving no pre-hospital critical care dispatch.
    In fact, the reality of hub decision-making on dispatch is that it must fit into a complex array of parallel tasks.
    These simultaneous demands can reach bewildering intensity once any of the teams has been tasked and is dispatched into the field. 
    This is because a tasked critical care team have a lot of practical and informational needs that the hub team have to meet.
    So there are continual rounds of planning and communication in the course of a tasking for which the hub team is central.
    These needs include the coordinating of active resources, as described, facilitating access to senior clinical consultation, logging critical care clinical milestones such as the administration of Controlled Drugs, providing meteorological and ground-crew status, and so on. 
    
    For the individual hub staff member, the work done at each step in the workflow is clearly interrelated. 
    Viewing a call builds on what is learned from the headline information.
    Interrogation is carried out to answer questions raised by the viewed call in relation to clinical status and likely indication for a critical care intervention.
    And any decision might need to be changed in light of new information. 
    Experts reflect this in their practice. 
    
    They view many calls with the expectation of quickly ruling out a need to interrogate. 
    These are frequently where the free text description in the headline listing doesn’t provide sufficient information to reassure that an emergency ambulance alone will be appropriate. 
    But, on the other hand, changed calls (which are indicated on their call system) may need to be re-examined. 
    So there will be a higher rate of viewing than there are distinct cases viewed in a given shift.

    The features of the informational landscape (the decision inputs) are increasingly complex as we move through the workflow steps. 
    Each prior decision is a decision to acquire more information. 
    And as emphasised by decision theory~\citep{howard2005InfluenceDiagrams}, informational influences will always impact those later decisions in a sequence. 

    While a high proportion of staff time is spent downstream, in interrogating a proportion of calls and in supporting and coordinating dispatched resources on the few calls that result in a tasking, the time spent making the upstream decision of whether to view a call or not can be variable and highly pressured.

    Having staff working in pairs undoubtedly improves their ability to monitor the call stack.
    Little time goes by when there isn't at least one person viewing the call stack - even if only briefly. 
    But at those times when both staff are interrogating calls or coordinating teams in the field, then there is a hiatus in monitoring, and a process that has to be undertaken to re-gain situational awareness of the stack as it has evolved while the team had no eyes on it.
    This period can be from a few minutes to over half an hour.
    Even if they are able to maintain continuous monitoring between them, there is always the need for each team member to return to the call list and re-gain their own situational awareness.
    The observation work reported in \cref{sssec:internal-and-external-interaction} shows how much this dominates the interactions of the staff.
    Any system modification that doesn't fit into this real context of situated use is of no value.
    Any decision support needs to be at once quickly accessible, easily reviewed and constantly unobtrusive.
    And for practical reasons of adoption, reliability and improvement, it also needs to be, at least in its early phases, amenable to efficient feedback processes.
    
    Slick and unobstructed re-assessment of the incoming call list isn't just a cognitive-visual challenge.
    The hub workflow we have outlined is one in which the user sometimes progresses by clicking in one system, sometimes progresses by using physical buttons in another system, breaks off to consult a third system and frequently returns to the first system to review, continue or retrace in order to re-review.
    This means the human agent is frequently involved in a complex array of interactions across four screens and multiple input devices.
    At the same time, they are undertaking frequent task-switching and significant mental caching.
    This places additional requirements on any assistive tool.
    It must be intuitively quick to access, to relate to a specific case and to interpret.
    At the same time, it must be unremarkable, unobtrusive and non-interruptive.

\subsubsection{Situated action - summary findings}
\label{sssec:situated-action-summary-findings}
We have examined the work of the two-person specialist team charged with how to find the less than one percent of emergency calls that can benefit from \textit{pre-hospital} critical care.
Our in-depth analysis of situated action at the critical care hub has allowed us to construct a detailed picture of how decision-making is accomplished.
We have established the key features of the workflow in the \textit{work-as-done}.
We found that there are multiple decisions on the way to a substantive dispatch decision.
Specifically, two triage steps are invariably undertaken in order to navigate the high volume of calls, filter (or sift) them down to a manageable number of candidate calls and then access the necessary information for a sufficiently-informed decision at a third step.
We have obtained a detailed picture of the trade-offs made concerning the time spent filtering cases and the time available to consider those filtered.

Our observation work revealed that the amount of interaction between hub staff is very high and indicates a great deal of joint work.
We see that the systems they use are not optimised for their work.
This is because they are designed for the wider ambulance service, which deals with the high volumes by segmentation of the calls among separate teams dedicated to each geographical region.

We have seen that situated decision-making is contingent and has no fixed thresholds.
Each step in the decision sequence is impacted by a number of evolving factors.
The individual call information is a focal influence which grows in detail through the sequence.
The context of actively monitored calls and those continually being added is a clinical and cognitive influence as it alters the assessment of where and how resources are likely to be disposed for the next minutes and hours.
Meanwhile, the wider disposition and availability of resources, which influences both information flows and immediate capacity, may alter in parallel with the sequence of decision steps.
Staff working at the hub maintain their own monitoring process, engage in recognition and action decisions and manage prioritisation, deconfliction and coordination of tasked teams in the field as well as providing critical care advice to regular ambulance crews and first responders.

Crucially, our ethnographic research has revealed that whilst a majority of interaction time is outward-facing -- making and interrogating external calls, the count of interaction types sees unsolicited situation awareness interactions dominate.
This indicates that the human team are constantly coordinating to maintain a shared mental model under conditions that are inherently volatile.
The high frequency of their interactions indicates the potential for rapid decay of situational awareness.
The economy of their utterances demonstrates a high level of mutual trust and reliance increasing the efficiency and robustness of their joint work.
This coordination work is undertaken in part to keep up with the evolving needs of individual cases.
But it is also to update the shared view of the disposition of resources now and in the near future precisely because this prediction influences the effective thresholds for decisions at all three key steps.

In terms of anticipating algorithmic assistance for the humans undertaking this work, there is a challenge surrounding establishing any equivalent of this shared situational awareness.
And a further challenge is in sourcing suitable training data.
Any algorithmic assistance system aiming to support the substantive dispatch decision would need to ingest a complex array of multi-modal data from sources not currently captured in accessible form.
Support in the second triage step would be able to ingest detailed call data, but it would need to support the trade-offs made by the hub staff in responding to the dynamics of demands and resources.

An algorithmic support at the first triage step has the challenge of having very little data per case, as well as not accessing anything close to a conventional ground truth.
But it has the advantage of having a clear and ready-made source of input data as well as an easily-navigated interface for support provision - the call list itself. 

The interface navigation question is worth dwelling on.
Finding an opportunity within the different display elements to insert `unremarkable' decision support~\citep{tolmie2002UnremarkableComputing,yang2019UnremarkableAIFitting} into this human experience is not simple.
Establishing whether such support can have a positive influence on users is even more challenging
Assistance needs to easily fit into and support a routine of attention-switching, scanning and assimilation.
For steps B and C there is a question-mark surrounding where in the workflow a supportive intervention could reasonably be designed.
While step A has a distinct advantage in its suitability for the efficient generation of expert feedback on regular samples of calls.
This is the closest thing to ground truth likely ever to be available, so consideration of how it is accumulated should be given prominence.

%% file: tbl_obs_interaction_types.tex
% !TEX root = ../outline.tex
\begin{table*}
	\centering
	\small
	\caption[Time between occurrences for different interaction groups]{Time between occurrences for different interaction groups}
	\begin{tabular}{l|l|l|r}
		\toprule

Int/Ext & Group & Description & Minutes between\\
        &       &             & occurrences (ave)\\

		\midrule

Int & Sit Aw Upd & Situational Awareness Update & 4\\
Ext & Op Com & Operational Communication & 7\\
Int & Sp Call & Specific Call Interaction & 18\\
Int & Oth Hub & Other Internal Hub Interaction & 20\\
Int & Non Tsk & Non-Task Interaction & 50\\
Int & Op Adv & Operational Advice & 55\\

		\midrule

\multicolumn{3}{c}{All interactions} & 2 \\
 % &  & All interactions & 2\\

		\bottomrule
	\end{tabular}
	\label{tbl:obs-interaction-types}

\end{table*}

%% file: tbl_obs_groups_by_shift_ave_per_hr.tex
    % \begin{figure}
    %     \centering
    %     \includegraphics[width=\linewidth]{tbl/obs_groups_by_shift_ave_per_hr.png}
    %     \caption{Caption to be added for obs-groups-by-shift-ave-per-hr}
    %     \label{tbl:obs-groups-by-shift-ave-per-hr}
    % \end{figure}

% !TEX root = ../outline.tex
\begin{table*}
	\centering
	\small
    \caption[Average count per hour interaction groups by shift]{Average count per hour of events in each interaction group in each of four shifts \\
    (see \cref{tbl:obs-interaction-types} for interaction Group descriptions).}
	\begin{tabular}{l|r|r|r|r|r}
		\toprule
            %% we add \,\,\,\, as a set of thin spaces to equalise the col widths without using p col types
            Interaction Group &  \,\,\,\,A Day & \,\,\,\,B Day  & C Night & D Night & Ave\\
		\midrule
            
  %           Group & 22SuD & 21SaD & 21SaN & 24TuN & Ave\\
		% \midrule
            Sit Aw Upd & 16.3 & 14.3 & 10.4 & 13.3 & 13.6\\
            Op Com$^a$ & 11.7 & 9.5 & 6.4 & 6.0 & 8.4\\
            Sp Call & 6.6 & 3.5 & 2.0 & 1.3 & 3.4\\
            Oth Hub & 3.7 & 5.6 & 2.0 & 0.7 & 3.0\\
            Non Tsk & 2.6 & 1.4 & 0.8 & - & 1.2\\
            Op Adv & 2.0 & 1.1 & 1.2 & - & 1.1\\
		\midrule
            Totals & 42.9 & 35.4 & 22.8 & 21.3 & 30.6\\
                    
		\bottomrule

	\end{tabular}
	\label{tbl:obs-gps-ave-hrly-four-shifts}
    \\
    \footnotesize a - external to the hub 
    \qquad\qquad\qquad
    \qquad\qquad\qquad
    \qquad\qquad\qquad
    \qquad\qquad\qquad  % spacing added to table footnote as hack, to appear left aligned

\end{table*}

%% This version of the inner table has zeros instead of dashes
% % Group & 22SuD & 21SaD & 21SaN & 24TuN & Ave\\

% Sit Aw Upd & 16.3 & 14.3 & 10.4 & 13.3 & 13.6\\
% Op Com & 11.7 & 9.5 & 6.4 & 6.0 & 8.4\\
% Sp Call & 6.6 & 3.5 & 2.0 & 1.3 & 3.4\\
% Oth Hub & 3.7 & 5.6 & 2.0 & 0.7 & 3.0\\
% Non Tsk & 2.6 & 1.4 & 0.8 & 0 & 1.2\\
% Op Adv & 2.0 & 1.1 & 1.2 & 0 & 1.1\\

%  & 42.9 & 35.4 & 22.8 & 21.3 & 30.6\\

%% file: tbl_obs_types_by_shift_ave_per_hr.tex
% !TEX root = ../outline.tex
\begin{table*}
	\centering
	\small
    \caption[Average count per hour interaction types by shift]{Average per hour of events of each interaction type in each of four shifts}
	\begin{tabular}{l|r|r|r|r|r}
		\toprule
            %% we add \,\,\,\, as a set of thin spaces to equalise the col widths without using p col types
            Interaction Type &  \,\,\,\,A Day & \,\,\,\,B Day  & C Night & D Night & Ave\\
		\midrule

            Sit Aw Upd - give & 14.9 & 8.6 & 7.6 & 10.0 & 10.3\\
            Sit Aw Upd - seek & 1.4 & 5.6 & 2.8 & 3.3 & 3.3\\            
		\midrule
            Op Com - ask/dir/get & 7.7 & 4.1 & 2.0 & 4.0 & 4.5\\
            Op Com - detail (clin/incidt) & 2.0 & 2.8 & 3.2 & 2.0 & 2.5\\
            Op Com - oth staff in room & 1.4 & 2.1 & 0.8 & - & 1.1\\
            Op Com - TCC & 0.6 & 0.5 & 0.4 & - & 0.4\\
		\midrule
            Sp Call - update, give & 4.6 & 1.3 & 1.2 & 1.3 & 2.1\\
            Sp Call - update, seek & 1.4 & 0.6 & 0.8 & - & 0.7\\
            Sp Call - ID, seek & 0.3 & 0.9 & - & - & 0.3\\
            Sp Call - ID, give/confirm & 0.3 & 0.8 & - & - & 0.3\\
		\midrule
            Oth Hub - affirm/negate & 1.7 & 3.3 & 1.2 & 0.7 & 1.7\\
            Oth Hub - assist, req/offer & 1.4 & 1.6 & 0.8 & - & 1.0\\
            Oth Hub - oth & 0.6 & 0.8 & - & - & 0.3\\
		\midrule
            Non Tsk - all & 2.6 & 1.4 & 0.8 & - & 1.2\\
		\midrule
            Op Adv - all & 2.0 & 1.1 & 1.2 & - & 1.1\\
		\midrule
            Totals & 42.9 & 35.4 & 22.8 & 21.3 & 30.6\\        
                    
		\bottomrule
	\end{tabular}
	\label{tbl:obs-typs-ave-hrly-four-shifts}

\end{table*}

%% file: incl_04_discussion.tex
% !TEX root = ../main.tex

\section{Discussion}
\label{sec:discussion}

    This paper sets out on a quest for ecological validity.    
    Cabitza and Zeitoun~\citep{cabitza2019ProofPuddingPraise} make a call to action on the need to demonstrate pragmatic and ecological validity.
    They outline a process that involves clinicians in evaluation against real-world end points.
    In the culture of `real-world validation', they advocate going beyond the initial technical question `can it work?' to the pragmatic question `does it work?' (beyond the bench) and the sustainability question `is it worth it?'~\citep[p5]{cabitza2019ProofPuddingPraise}.
    
    In a related but distinct call, Zajac et al~\citep{zajac2023ClinicianFacingAIWild} advocate for an earlier shift -- in the development process itself.
    They identify a problem that they call `late realisation', meaning that extensive data gathering, processing and modelling work consumes the major share of development time. 
    And the common practice is, they argue, that consideration of clinician-facing evaluation is left to the later part of any AI project.
    To tackle this `late realisation', the developer community is urged to engage in extended interdisciplinary work in order for the needs of situated workflows to penetrate into earlier development processes.
    At the same time, the authors endorse `near-live' experimentation, to ensure meaningful evaluation. 

    The call of these writers, then, is for ecological validity and the avoidance of late realisation.
    Time invested in the study of situated action is a practical response to both these calls.
    Our immersion and observation work ensures that the researcher appreciates the real-world context at a depth that tests the points of commonality required for productive interdisciplinary work.
    
    Contextual inquiry and appreciation enables developers to make better choices at the outset of the system design process.
    They can push for interdisciplinary processes to begin earlier in the development timeline (a software development practice they already know as `shift left') and they can engage with clinical users on terms that both sides can understand.
    At the same time, the points of commonality enable clinical leaders and frontline practitioners to see what situated evaluation involves for their workplaces - in terms that non-clinicians can readily grasp.
    All stakeholders can comprehend what it will mean to approach `near-live' experimentation.
    That is, where algorithmic contributions fit in the workflow and what humans are doing at that point.

    Making use of this shared perspective allows us to look at a proposed deployment and understand what we need to pay attention to in order to be confident it will work sufficiently well, and will continue to work sufficiently well in the future in a real-world setting, for it to be worthwhile. 
    And once translation is accomplished, we can monitor it more easily because the elements of the situated workflow are understood.

    We provide substance in the form of weight and texture with detailed observational research. Relatively simple ethnographic techniques locate and uncover the kinds of opportunities for, and barriers to, algorithmic assistance in real human workflows.

    Decision-making is always situated in some context that is significant -- other decisions, other actions, processes that change over time.
    Close readings of sociologically informed texts by, among others, Lucy Suchman~\citep{suchman1985plans} and Paul Dourish~\citep{dourish2001WhereActionFoundations} were influential in the design and conduct of the observational work. 
    Ethnographic methods led the researcher to uncover the daily enacted practices of hub staff that informs our understanding of where and how assistance can be effective as well as where it would be more likely to fail.
    Our extended informal interviews and discussions during and around the fixtures of the working day and in internal meetings exposed details that were not available in management descriptions nor described in documented procedures.
    What this work revealed was crucial information on which step in the situated workflow to focus on.
    
    Witnessing and recording the work-as-done as well as the accounts by clinicians of their own decision-making practice allowed us to more carefully define the task to be supported and the competing demands to be navigated.
    We were able to see the way in which decisions are influenced - not just by considering decisions individually - but by how they interact, by how they influence each other - the whole decision-making workflow.
    
    Clinicians' own descriptions of what information they take into account when making decisions at the different steps of their workflow provided a rich understanding of their situated decision-making. And this understanding was enhanced by their own accounting of the mental processes they used when fitting their own on-scene experience of the field to call interrogations and documented call headlines at the hub.

    At the outset, we had placed significant weight on the problem of organisational friction in our estimation of feasible development.
    In particular we estimated that insurmountable challenges would be involved in having an algorithm ingest large quantities of back-end data that weren't readily exposed to any application interface. 
    But it was the strong expression of clinicians' own preference for assistance at the first triage step that led to a decision to place exclusive focus on that part of the workflow for subsequent experimental exploration.
    In summary, we took an approach that leveraged a significant understanding of the real decision-making context to ensure that development and evaluation work would align with situated action.

%% file: incl_05_conclusions.tex
% !TEX root = ../main.tex

\section{Conclusions}
\label{sec:conclusions}

    The contribution here takes the form of observational field work that serves to demonstrate how designing for situated decision-making requires a granular understanding of the situated action of decision-makers themselves.

    This work draws on the significant material contributed to HCI by researchers skilled in sociological techniques.
    The ethnography is neither complex nor particularly extensive.
    But the essential requirement is that the observation is attentive to the specific detail of our chosen work environment.
    And this requirement is met by ensuring that the time is spent sufficient to embed the observer in the daily practices of the staff to be supported in the decision task.
    The resulting insight is able to inform subsequent design work - both in terms of where support is provided and the form it takes.
    
    We draw not only practical but also methodological conclusions from our work observing staff at the critical care hub.
    Undertaking such work informs and enriches our understanding of the decision context and can enable appropriately designed interventions when it comes to providing assistive tools.
    As a result, the focus of the technological intervention will be the early decision to view which is only one part of a multi-step triage process.
    And the intervention itself will be an unobtrusive simple score, displayed within the existing call list that is already a central part of the workflow.

    Methodologically, we see that our emphasis on participatory design and a deep analysis of situated work-as-done rather than work-as-documented or work-as-imagined is what makes it possible to gain the necessary depth of understanding of the task and its context. 
    Understanding the task in context (the situated task) is what enables a relatively modest intervention to receive engagement and praise from practitioners and produce insight into the likely effect of algorithmic assistance.

    We contend that these approaches can be applied in a wide variety of situated decision-making contexts.
    While we have confined ourselves to a specific clinical decision-making context, there is every reason to suppose that our approach will apply with equal value in other decision-making contexts.
    This is because we have focused primarily on the sociotechnical elements in drawing our conclusions and not the domain-specific elements of the decision-context. 

    We assert that our conclusions from the study of situated action are inherently generalisable.
    Studying situated action applies to any aspect of human decision making work.
    Indeed, it is the only way in which we can hope to find out whether and how techniques from one context will be of any practical value in another - or what modifications may need to be made to an approach, a design or a workflow.
            
    The prospects for meaningful and useful algorithmic development work are good, provided we allow efforts to be directed by the needs of the situated decision-making task.

%% file: incl_99_declarations.tex
% !TEX root = ../main.tex

\section*{Acknowledgements}
We are grateful to the staff of the EMRTS Critical Care Hub for their patience and feedback. Specific thanks are also due to Mark Winter, Operations Director at EMRTS, Cymru, Kate Blackmore, area manager for WAST CCC, Cwmbran and Robert Brunnock, service manager for EMS Coordination, WAST.

\section*{Declarations}

\subsection*{Funding}
B. Wilson gratefully acknowledges support for this work from the UK Engineering and Physical Sciences Research Council grant EP/S021892/1, from EMRTS Cymru and from the European Union.
M. Roach acknowledges support for this work from the European Union.

Funded by the European Union. Views and opinions expressed are however those of the author(s) only and do not necessarily reflect those of the European Union or the European Health and Digital Executive Agency (HaDEA). Neither the European Union nor HaDEA can be held responsible for them. Grant Agreement no. 101120763 - TANGO.

\section*{Ethics statement} 
Ethical approval for this study was granted by the UK Health Research Authority and Health and Care Research Wales under IRAS project IDs 312173 (immersion) \& 316886 (observations) in addition to Confidentiality Advisory Group approval under application 24CAG0106 (observations). Patient and public involvement and engagement (PPIE) consisted of attendance, presentation and discussion at PPIE meetings. Feedback, suggesting accessible information for the public be made available, led to a dedicated page appearing on the EMRTS website. Participants (EMRTS staff) gave informed consent to participate in the study before taking part.

\section*{Contribution statement}
\textbf{BW}: Conceptualisation, Project administration, Methodology, Investigation, Data curation, Formal analysis, Visualisation, Writing - original draft, Writing - review \& editing. 
\textbf{MR}: Resources, Supervision, Funding acquisition, Writing - review \& editing.
\textbf{GB}: Resources, Project administration.
\textbf{CC}: Resources.
\textbf{DR}: Resources, Supervision, Funding acquisition.

\section*{Competing interests}
The authors declare no competing interests.

\section*{Open Access}  
For the purpose of Open Access, the author has applied a CC BY licence to any Author Accepted Manuscript (AAM) version arising from this submission.

\section*{Use of Generative AI}
The authors made no use of generative AI in the preparation of the text of this manuscript. 
% Generative AI was used to create a first draft of figure descriptions. 
% Each was checked and edited manually for readability. 
Generative AI was used to create the sketches from field photographs in order to avoid individuals being identifiable in the figures.  
Spelling and grammar check tools were in use during authoring.